\documentclass[11pt]{article}

\usepackage{amsthm,amssymb,amsmath}
\usepackage{amsmath,amsthm,amssymb,amsfonts}
\usepackage[english]{babel}
\usepackage{hyperref}
\usepackage{natbib}
\usepackage{multirow}
\usepackage[
letterpaper,
left=1.21in,
right=1.21in,
top=1in,
bottom=1in
]{geometry}
\usepackage{changepage}
\usepackage{xcolor}
\usepackage{geometry}
\usepackage{graphicx}
\usepackage[capposition=top]{floatrow}
\usepackage{subcaption}
\usepackage{setspace}
\newcounter{subsubsubsection}[subsubsection]

\usepackage{authblk}

\hypersetup{
	colorlinks,
	linkcolor={red!50!black},
	citecolor={blue!50!black},
	urlcolor={blue!80!black}
}
\newtheorem{theorem}{Theorem}
\newtheorem{proposition}{Proposition}
\newtheorem{corollary}{Corollary}
\newtheorem{lemma}{Lemma}
\newtheorem{assumption}{Assumption}
\newtheoremstyle{style2}{4pt}{4pt}{}{}{\bfseries}{.}{ }{} 
\theoremstyle{style2}

\numberwithin{lemma}{section}
\numberwithin{proposition}{section}
\numberwithin{remark}{section}

\title{Inference on Quantile Treatment Effects with Small Treatment Groups\thanks{We would like to thank Fred Finan, Michael Jansson, and Todd Messer for useful comments and discussions. The usual disclaimer applies.}}
\author{Marina Dias\thanks{e-mail: mvbdias@amazon.com. Address: Amazon Inc., 1800 South Bell Street, Arlington VA. Note that this work was done prior to joining Amazon.}~~and Demian Pouzo\thanks{Corresponding author. e-mail: dpouzo@berkeley.edu. Address: UC Berkeley, Dept. of Economics; 530
			Evans Hall, Berkeley CA 94720 USA.}}

\begin{document}
	\maketitle
	\begin{abstract}
We develop inference for conditional quantile treatment effects that allows treatment groups to remain small as the overall sample grows. Such groups arise naturally in multivalued treatment models. Our procedure combines estimation of a conditional quantile function from a large reference group with an exact conditional Bernoulli representation of quantile indicators. We establish asymptotic size control and derive a power bound allowing treatment-group sizes to remain bounded. The framework accommodates nonseparable outcome functions, heteroskedasticity, and flexible error distributions across treatment profiles. We illustrate its scope through dynamic treatment models, nonseparable quantile models, and difference-in-differences designs with staggered adoption. Monte Carlo experiments assess finite-sample performance, and an application to temperature shocks and GDP growth illustrates inference on effects associated with distinct treatment histories.
	\end{abstract}

\section{Introduction}

Many applications in economics require accounting for treatment effect heterogeneity across individuals, firms, or regions and across the timing and sequence of treatment. In panel data, outcomes may depend on the entire treatment profile, with the effect of a treatment episode varying according to treatments received in earlier periods. The sequence of past and current treatment realizations thus defines a multivalued treatment. Even when treatment is binary at each date, the number of possible profiles can be large, creating many potentially distinct treatment effects.

Accommodating this heterogeneity requires distinguishing treatment groups that may contain few observations, even when the overall sample is large.  Pooling observations across profiles can increase precision, but requires restrictions that may be violated when treatment effects exhibit cumulative effects or adaptation to repeated exposure, leading to model misspecification. Small treatment groups can therefore arise naturally when allowing treatment effects to differ across profiles. In these settings, a large overall sample does not by itself justify conventional inferential procedures. The number of observations informative about a particular treatment effect may remain small, thus rendering standard inferential procedures untrustworthy or even invalid.

This paper contributes to the literature by proposing a method for inference on conditional quantile treatment effects that remains asymptotically valid when the number of observations in some treatment groups stays small as the overall sample grows. Conditional quantile effects capture distributional consequences of treatment that average effects can conceal (e.g.,  \citet{abadie2002instrumental,chernozhukov2005iv}). In particular, they distinguish shifts near the center of the conditional outcome distribution from changes in its tails, thereby offering a perspective complementary to average effects on the distributional consequences of treatment. To study these effects across treatment histories, we develop the method in a dynamic potential outcomes framework in which a structural outcome function determines the current potential outcome from the treatment profile, past potential outcomes, observed characteristics, and an unobserved rank. What we think is a salient feature of our framework is its generality; it accommodates multivalued treatments and allows the outcome function to be nonseparable and flexible in functional form, with heteroskedastic errors and flexible error distributions across treatment profiles. We test whether the structural outcome functions for a user-specified set of treatment profiles equal the control-profile function.

To establish the causal interpretation of this hypothesis, we show that it is equivalent to the absence of conditional quantile treatment effects for the selected profiles relative to the control profile. Differences in the function across treatment profiles therefore represent conditional quantile treatment effects, holding past outcomes and observed characteristics fixed.

The key identifying assumption is a quantile analogue of selection on observables: conditional on past potential outcomes and observed characteristics, treatment selection preserves the quantile of interest of the unobserved rank, while allowing other features of its distribution to vary across treatment profiles. The restriction accommodates a range of settings, including randomized experiments, observational models satisfying conditional independence, and dynamic treatment assignments satisfying suitable restrictions on selection given past outcomes and observed characteristics (see, e.g., \citet{rosenbaum2010design}, \citet{imbens2000role}, and
\citet{firpo2007efficient}). It also connects to quantile control-function approaches, in which conditioning on a suitable control variable accounts for endogeneity and restores the relevant conditional quantile restriction; see e.g. \cite{lee2007endogeneity}. In the difference-in-differences model with staggered adoption developed below, it takes the form of a quantile analogue of the conditional parallel-trends assumption commonly used to identify mean treatment effects (e.g. \cite{abadie2005}).

Our method builds on the conditional quantile moment restriction implied by the identifying assumptions. Under the null, the reference-profile quantile function satisfies this restriction in each treatment group under consideration. We first estimate this function using the large control/reference group; any estimator satisfying an approximation condition can be used, but we use a sieve-based conditional moment estimator following \cite{chen2009efficient,chen2012estimation}. We then construct empirical moments in each treated group by evaluating quantile residuals at the estimated reference quantile function and interacting them with basis functions of past outcomes. The test statistic measures the distance of these empirical moment vectors from zero and rejects when that distance is sufficiently large. Its construction is thus analogous to a specification test based on a semi/nonparametric GMM criterion.

The novelty relative to this literature lies in approximating the distribution of the test statistic when the sample sizes underlying its moments may grow at different rates or remain bounded. In particular, a Gaussian approximation to the moments of a treatment group need not be valid. Building on \cite{chernozhukov2009finite}, we instead exploit an exact conditional representation: under the null, conditional on past outcomes, observed covariates, and treatment profiles, the indicators evaluated at the true reference quantile function are independent Bernoulli variables with success probability equal to the quantile index. Consequently, the joint conditional distribution of the oracle moment vectors, and hence of the oracle statistic, can be simulated using independent Bernoulli draws while holding the observed histories and treatment profiles fixed. We provide sufficient conditions on the  estimator under which replacing the true reference function with its estimator changes the statistic by an asymptotically negligible amount. A vanishing adjustment to the simulated critical value accounts for this estimation error. We prove asymptotic size control and derive an explicit lower bound on power, both allowing treatment-group sizes to remain bounded.

These results permit valid testing without consistently estimating a separate quantile function for each treatment profile. In specifications that encode treatment effects in a finite-dimensional parameter, test inversion also yields confidence regions with asymptotically valid coverage. Validity, however, does not eliminate the information constraints imposed by small treatment groups: a growing reference group improves estimation of the benchmark but need not increase the information available about a sparsely observed treatment effect. Our power bound quantifies this relationship by relating detectability to the magnitude of the quantile effect, the conditional probability mass near the tested quantile, and the treatment-group size, with an explicit Berry–Esseen remainder. When a group grows, the bound recovers the familiar inverse-square-root scale governed by that group’s size under suitable conditions. When a group remains bounded, the bound continues to apply and provides sufficient conditions under which the test can detect an effect with a given probability. Our Monte Carlo experiments illustrate how these factors affect the ability to detect treatment effects.

We apply our framework to three examples that illustrate distinct benefits of the method. The first generalizes a canonical dynamic treatment model by allowing interactions across treatment periods, heteroskedasticity, and conditional error distributions that differ across treatment profiles. While we impose no functional form restrictions on this conditional distribution, we show that the estimation problem remains parametric, thus yielding root-$n$ estimation of the reference/control quantile function. The second illustrates inference in nonseparable structural models, where treatment effects may vary with observed characteristics, past outcomes, and latent ranks, even when some treatment groups remain small. The third considers difference-in-differences with staggered adoption, providing a testing procedure for quantile and average treatment effects when some adoption cohorts contain only finitely many observations under a quantile analogue of parallel trends. We show that in the baseline case considered in the literature, the quantile treatment effect coincides with the average treatment effect. The framework, however, also accommodates nonseparable extensions of the baseline additive model, allowing quantile effects to differ from average effects. In the first and third examples, test inversion yields confidence regions for the treatment-effect parameters. We provide sufficient conditions for implementing the procedure in each setting.

\paragraph{Related Literature.} Our paper builds on and contributes to several strands of literature. First, we contribute to the literature on inference with few treated observations and many controls.  In difference-in-differences settings, \cite{conley2011inference} and \cite{ferman2019inference} use information from a large collection of control groups to conduct inference with few treated groups, with the latter explicitly allowing for certain forms of heteroskedasticity. Outside the difference-in-differences framework, \cite{ferman2021} studies matching estimators with few treated and many control observations and develops inference based on approximate symmetry. Our approach shares the use of a large control sample but differs in the restriction that permits inference in the treated groups. This allows for heteroskedasticity and flexible error distributions across treatment profiles without requiring equality of their full distributions or symmetry of the errors. It also accommodates nonseparable outcome functions and semiparametric or nonparametric specifications.

Second, a related literature develops inference with a small number of clusters; see \cite{donald2007inference,cameron2008cluster,hagemann2020inference,hagemann_permutation_2023,Canay2017randomization,CanaySantosShaikh}. The results in these papers essentially hinge on the assumption that the distribution of residuals is, at least asymptotically, symmetric. This assumption can be justified with large within-cluster samples. Of particular relevance, \cite{hagemann2023inference} develops inference on quantile processes, including quantile treatment effects, with a finite number of large heterogeneous clusters. Our framework instead permits the number of observations within a treatment group itself to remain bounded.

Third, the literature on estimation and inference of conditional quantile models (e.g., \cite{chen2007large,chen2009efficient,chen2012estimation,horowitz2007nonparametric,gagliardini2012nonparametric}) supplies methods for estimating the reference-profile function from the large reference group. The additional challenge in our setting is that the empirical moments entering the test can have different rates of convergence, and some need not concentrate as the overall sample grows. We address this challenge by combining reference-function estimation with the finite-sample quantile inference approach of \cite{chernozhukov2009finite}.

Finally, our paper relates to work on staggered adoption in difference-in-differences designs (e.g., \cite{ChaiseMartin2020,callaway2018difference,SUN2020}) and to the literature on distributional and quantile effects in such designs (e.g. \cite{atheyimbens2006,callawayli2019}). Our contribution addresses the inferential difficulties that arise when adoption cohorts are small. In the baseline model considered by the literature, the quantile treatment effects coincide with the average treatment effects (ATE) and average treatment effect on the treated. Our method, however, can handle extensions of this baseline model to allow for nonseparabilities and more flexible functional forms, in which the equivalence between QTE and ATE is broken.

\section{Illustrative Example and Preview of the Results}
\label{exa:Canon} \label{sec:illustrative}

We now present a simple canonical model, which illustrates our testing procedure. For each unit $i$ and time $t$, let $Y_{i,t}$ be the economic outcome of interest and let $D_{i,t}$ be a binary variable that encodes the treatment status for this unit at time $t$.  We postulate that \footnote{Here and throughout, for any random variable, $A^{t} : = (A_{1},...,A_{t})$. Also, additional variables such as exogenous covariates and region-time fixed effects are omitted to simplify the exposition.}
	\begin{align}\label{eqn:Canon-1}
		Y_{i,t} = \sum_{d^{t} \in \{0,1\}^{t}} \theta_{d^{t}} 1\{D^{t}_{i} = d^{t}\} + \gamma^{\prime} Y^{t-1}_{i} + \sigma (Y^{t-1}_{i},D^{t}_{i})\epsilon_{i,t},\quad \sigma(Y^{t-1}_{i},D^{t}_{i}) > 0. 
	\end{align}

This model encompasses specifications used in a wide range of economic applications. Related models have been employed to study the effects of weather and climate shocks on economic outcomes (\cite{dell2012temperature,dell2014we,burke2015climate}), economic shocks on conflict (\cite{miguel2004economic,bazzi2014economic}), democratization on economic growth (\cite{acemoglu2019democracy}), and investments in school facilities on educational and housing outcomes (\cite{cellini2010value}).

Expression \ref{eqn:Canon-1} assigns a separate coefficient, $\theta_{d^{t}}$, to each possible treatment history $d^{t}$, allowing the effect of treatment at one date to depend arbitrarily on treatment realizations at other dates. The familiar additive distributed-$L$ lag model is nested as the special case in which $\theta_{d^{t}} = \kappa + \sum_{s=0}^{L}\beta_s d_{t-s}$ for all $L \leq t-1$. Under this restriction, observations with different treatment histories can be pooled to estimate the common coefficients $(\beta_s)_{s=0}^{L}$, potentially producing substantial gains in precision. These gains, however, rely entirely on additivity across treatment dates, ruling out interactions such as cumulative, compounding, and adaptation effects. For example, the restriction requires the effect of two consecutive weather shocks to equal the sum of the effects of the two shocks considered separately. Such restrictions may be difficult to justify in many applications. If these restrictions are incorrect, pooling across treatment histories can conceal economically important heterogeneity and lead to misleading conclusions about the effects of particular treatment sequences. The resulting small-group problem is precisely the cost of avoiding extrapolation across potentially heterogeneous treatment histories, and our method provides valid inference without relying on such extrapolation.

Our point of departure is to treat each history $d^{t}\in\{0,1\}^{t}$ as a distinct treatment, with an associated potential outcome $Y_t(d^{t})$. This formulation allows treatment effects to depend arbitrarily on the timing and sequence of past treatments, but it generates $2^t$ treatment profiles. Consequently, even with a moderately large overall sample, some treatment-profile groups may contain very few observations.

To illustrate this problem, we use data from \citet{dell2012temperature} and define $D_{i,s}=1$ when country $i$ experiences an increase in temperature of more than one degree Celsius between periods $s-1$ and $s$. Table \ref{tab:DJE.Groups} reports the number of countries associated with each five-year treatment profile in three different periods. The control profile contains between 93 and 113 countries, whereas every noncontrol profile contains fewer than 20 countries and most contain fewer than 10. Because Expression \ref{eqn:Canon-1} assigns a separate parameter $\theta_{d^t}$ to each profile, conventional large-sample approximations may be unreliable for inference on the sparsely populated profiles. Our objective is to provide valid inference when the control group is large but the number of observations in the treatment groups may remain small, even asymptotically.

	\begin{table}[h]
		\begin{center}
\begin{tabular}{lrrr}
  \hline
Profile & N 1960-1964 & N 1980-1984 & N 1990-1994 \\ 
  \hline
00000 & 100 & 113 &  93 \\ 
  00001 &   7 &  &  17 \\ 
  00010 &   1 &   5 &   3 \\ 
  01000 &   4 &   1 &  \\ 
  01001 &   1 &  &  \\ 
  10000 &   7 &   1 &   6 \\ 
  10100 &   4 &  &  \\ 
  00100 &  &   3 &   1 \\ 
  01010 &  &   1 &  \\ 
  00011 &  &  &   1 \\ 
  00101 &  &  &   2 \\ 
  10001 &  &  &   1 \\ 
   \hline
\end{tabular}

			\caption{Number of countries for each treatment profile with $L=4$ and for different periods $t \in \{ 1964, 1984, 1994 \}$. Empty groups mean no countries for that profile. Data: Panel of 124 countries taken from \cite{dell2012temperature}.}
			\label{tab:DJE.Groups}
		\end{center}
	\end{table}

\paragraph{Preview of the Testing Procedure.} To preview our testing procedure, we specialize expression \ref{eqn:Canon-1} to a five-period treatment history and suppress past outcomes and other covariates. We consider
\begin{equation}\label{eqn:Canon-1.a}
	Y_i
	=
	\sum_{d\in\{0,1\}^{5}}
	\theta_d 1\{D_i^5=d\}
	+
	\sigma(D_i^5)\epsilon_i,
\end{equation}
where $\sigma(d)>0$ for every $d\in\{0,1\}^5$, $\epsilon_i$ is independent of $D_i^5$, has no atoms, and $\operatorname{Med}(\epsilon_i)=0$. As will become apparent below, this quantile restriction will be crucial for our analysis --- on the other hand, the independence restrictions are imposed only to make the main idea transparent; Section \ref{sec:setup} replaces them with a weaker conditional quantile restriction.

Suppose we are interested in testing the hypothesis that the median causal effect of a certain treatment profile, say the combination $\delta_{1}: = (0,0,0,1,0)$, is equal to that of the ``control" treatment given by $\delta_{0} : = (0,0,...,0)$; i.e., $Med(Y(\delta_{1})) = Med(Y(\delta_{0}))$.

It is easy to see from expression \ref{eqn:Canon-1.a} and the exogeneity of the treatment profile that this hypothesis translates to
\begin{align*}
 \operatorname{H}_0 ~:~	\theta_{\delta_1} - \theta_{\delta_0} = 0
\end{align*}

 Unfortunately, to test this hypothesis we cannot rely on direct estimation of the parameter $\theta_{\delta_1}$ due to the potentially small number of treated units for $\delta_1$ (recall Table \ref{tab:DJE.Groups} above). 
 
 To sidestep these issues, we exploit the median restriction of our model as follows. Suppose for now that $\theta_{\delta_0}$ is known. Then under $H_0$,
 \[
 \sum_{i:D_i^5=\delta_1}1\{Y_i\leq\theta_{\delta_0}\}
 \,\bigg|\,
 (D_i^5)_{i=1}^n
 \sim
 \operatorname{Binomial}\left(N_{\delta_1},\frac{1}{2}\right),
 \]
 where $N_{\delta_1}:=\sum_{i=1}^n1\{D_i^5=\delta_1\}$. Based on this observation, we can construct the (oracle) test statistic:
 \[
 Q_n(\theta_{\delta_0})
 :=
 \left|
 \frac{1}{N_{\delta_1}}
 \sum_{i=1}^{n}
 1\{D_i^5=\delta_1\}
 \left(
 1\{Y_i\leq\theta_{\delta_0}\}
 -\frac{1}{2}
 \right)
 \right|.
 \]
 Under $H_0$, conditional on $(D_i^5)_{i=1}^n$, $ Q_n(\theta_{\delta_0})$ has the same distribution as
$\left|
 B_{N_{\delta_1}}/N_{\delta_1}
 -0.5
 \right|$, where $B_{N_{\delta_1}}
 \sim
 \operatorname{Binomial}
 \left(N_{\delta_1},\frac{1}{2}\right)$. Large values of $ Q_n(\theta_{\delta_0})$ therefore provide evidence against $H_0$. In particular, a level-$\alpha$ test can be constructed using the conditional critical value
 \[
 c_{\alpha}(N_{\delta_1})
 :=
 \inf
 \left\{
 c\geq 0:
 \Pr\left(
 \left|
 \frac{B_{N_{\delta_1}}}{N_{\delta_1}}
 -\frac{1}{2}
 \right|
 \leq c
 \right)
 \geq 1-\alpha
 \right\},
 \]
 and rejecting $H_0$ when $Q_n(\theta_{\delta_0})>c_{\alpha}(N_{\delta_1})$. Because the critical value is computed from the binomial distribution conditional on $N_{\delta_1}$, this oracle test does not require the number of observations with profile $\delta_1$ to grow with the overall sample size.
 
 In practice, $\theta_{\delta_0}$ is typically unknown. However, because the number of observations with the control profile $\delta_0$ is large, we can estimate it consistently using the control-group sample median, denoted by $\widehat{\theta}_{\delta_0}$ --- this estimator is constructed exploiting the quantile restriction $Med(Y|D^{5}=\delta_0) = \theta_{\delta_0}$. The feasible statistic is
 \[
 \widehat{Q}_n
 :=
 \left|
 \frac{1}{N_{\delta_1}}
 \sum_{i=1}^{n}
 1\{D_i^5=\delta_1\}
 \left(
 1\{Y_i\leq\widehat{\theta}_{\delta_0}\}
 -\frac{1}{2}
 \right)
 \right|.
 \]
 As the size of the control group increases,
 $\widehat{\theta}_{\delta_0}$ converges to $\theta_{\delta_0}$ and
 $\widehat{Q}_n$ is asymptotically equivalent to $ Q_n(\theta_{\delta_0})$. Consequently, the conditional binomial distribution above also provides an asymptotically valid reference distribution for the feasible test, even when $N_{\delta_1}$ remains small. Sections \ref{sec:setup} and \ref{sec:main} formalize this argument in the general model.

Several features of this procedure are worth emphasizing.
\begin{enumerate}
	\item The argument requires the size of the control group to diverge, but it does not require the size of the treatment group, $N_{\delta_1}$, to diverge. 
	
	\item The median is used only for simplicity. The same argument applies to any quantile $\tau\in(0,1)$. Moreover, treatment profiles need not be unconditionally exogenous; see Assumption \ref{ass:U-DGP} below.
	
	\item The oracle reference distribution is binomial and therefore does not require homoskedasticity or a parametric specification of the residual distribution. The feasible test inherits this robustness asymptotically as the control-group estimator becomes consistent.
	
	\item The general framework allows treatment effects to depend on observed covariates and past outcomes and accommodates nonseparable and nonparametric outcome functions. Following \citet{chernozhukov2005iv} and \citet{han2018identification}, we represent potential outcomes as
	\begin{equation}
		Y_t(d^t)
		=
		\mu_t\left(
		Y^{t-1}(d^{t-1}),
		X^t,
		d^t,
		U_t(d^t)
		\right),
		\qquad
		d^t\in\{0,1\}^t,
	\end{equation}
	where $U_t(d^t)$ is an unobserved rank variable. For the illustrative model, if $F_\epsilon$ denotes the CDF of $\epsilon_t(d^t)$, then $U_t(d^t)
	=
	F_\epsilon\bigl(\epsilon_t(d^t)\bigr)$ 	and
	\[
	\mu_t\left(
	y^{t-1},
	x^t,
	d^t,
	u
	\right)
	=
	\theta_{d^t}
	+
	\sigma(d^t)F_\epsilon^{-1}(u).
	\]
	Section \ref{sec:setup} develops this representation and its identifying restrictions.
\end{enumerate}

\section{Setup}\label{sec:setup}

We consider a panel with a fixed number of periods
$\mathcal{T}:=\{1,\ldots,T\}$. Units are indexed by
$i\in\{1,\ldots,n\}$. For each unit and period, we observe an
outcome $Y_{i,t}\in\mathcal{Y}\subseteq\mathbb{R}$, a binary treatment
$D_{i,t}\in\{0,1\}$, and a vector of covariates
$X_{i,t}\in\mathcal{X}\subseteq\mathbb{R}^{q}$. We use $D_i^t:=(D_{i,1},\ldots,D_{i,t})$ and $X_i^t:=(X_{i,1},\ldots,X_{i,t})$ to denote their histories through period $t$. 

For every treatment history $d^t=(d_1,\ldots,d_t)\in\{0,1\}^t$, let
$Y_{i,t}(d^t)$ denote the potential outcome of unit $i$ in period $t$
under treatment history $d^t$. For $s\leq t$, let
$d^s:=(d_1,\ldots,d_s)$ and define the corresponding potential-outcome
history by $Y_i^t(d^t)
:=
\bigl(
Y_{i,1}(d^1),\ldots,Y_{i,t}(d^t)
\bigr)$. 

We allow treatment assignment to depend on observed histories. In
particular, treatment may be generated according to
\begin{equation}\label{eqn:TE-process}
	D_{i,t}
	=
	\delta_t\left(
	Y_i^{t-1},
	X_i^t,
	D_i^{t-1},
	V^{t}_{i}
	\right),
\end{equation}
where $V_{i}^{t}$ is an unobserved treatment-assignment profile shock. Expression
\eqref{eqn:TE-process} permits current treatment to depend on
past outcomes, past treatments, and current and past observed
covariates.

Potential outcomes satisfy the structural representation
\begin{equation}\label{eqn:PO-process}
	Y_{i,t}(d^t)
	=
	\mu_t\left(
	Y_i^{t-1}(d^{t-1}),
	X_i^t,
	d^t,
	U_{i,t}(d^t)
	\right),
	\qquad
	d^t\in\{0,1\}^t,
\end{equation}
where $U_{i,t}(d^t)\in[0,1]$ is an unobserved rank variable. The
outcome function $\mu_t$ determines how the treatment history,
observed characteristics, past potential outcomes, and unobserved rank
jointly determine the current potential outcome.

Under the assumptions below, $\mu_t\left(y^{t-1},x^t,d^t,\tau\right)$ 
is the conditional $\tau$-quantile of the potential outcome
$Y_t(d^t)$ for units with past potential-outcome history
$Y^{t-1}(d^{t-1})=y^{t-1}$ and observed characteristics
$X^t=x^t$. Consequently, for fixed $(y^{t-1},x^t,\tau)$, the mapping
\[
d^t
\mapsto
\mu_t\left(y^{t-1},x^t,d^t,\tau\right)
\]
has a causal interpretation in terms of conditional structural
quantile treatment effects. We formalize this interpretation below.

\paragraph{Assumptions.} We impose the following conditions. Here and throughout, let $P_{d^{t},x^{t}}(\cdot)$ denote $P(Y^{t-1} \in \cdot \mid D^{t} = d^{t} ,X^{t} = x^{t} )$.

\begin{assumption}[Sampling]\label{ass:DGP}
	For each $n$, $
	\bigl(
	Y_{i,t}(d^t),
	U_{i,t}(d^t)
	\bigr)_{t\in\mathcal{T},\,d^t\in\{0,1\}^t},
	X_i^T,
	D_i^T,
	V_i^T)$, $i=1,\ldots,n$ are independently drawn from $P$, such that expressions \ref{eqn:TE-process} and \ref{eqn:PO-process} hold and $D_{0}$ and $Y_{0}(.)$ are assumed to be exogenous; $\mathcal X^{T}$ is finite; and $P_{d^{t},x^{t}} << P_{0^{t},x^{t}} $ a.s.
\end{assumption}


\begin{assumption}[Monotonicity]\label{ass:monotonicity}
	For every $t\in\mathcal{T}$ and every
	$(y^{t-1},x^t,d^t)$, the mapping
	\[
	u
	\mapsto
	\mu_t(y^{t-1},x^t,d^t,u)
	\]
	is continuous and strictly increasing.
\end{assumption}

\begin{assumption}[Conditional rank restriction at $\tau$]
	\label{ass:U-DGP}
	There exists a $\tau \in (0,1)$ such that for every $t\in\mathcal{T}$ and every
	$d^t\in\{0,1\}^t$,
	\begin{equation}\label{eqn:conditional-rank}
		P \left(
		U_{i,t}(d^t)\leq\tau
		\,\middle|\,
		Y_i^{t-1}(d^{t-1}),
		X_i^t,
		D_i^t
		\right)
		=
		\tau
		\qquad \text{a.s.}
	\end{equation}
	Moreover, the conditional distribution function of
	$U_{i,t}(d^t)$ is continuous and strictly increasing in a
	neighborhood of $\tau$.
\end{assumption}

We discuss these assumptions in Section \ref{sec:Discussion.Assumptions} below. We now link the outcome function $\mu_{t}$ to a conditional quantile function which allows us to cast the null hypothesis in terms of QTEs. 

\paragraph{Conditional quantile treatment effects.}For any $d^t\in\{0,1\}^t$, let
\[
q_{\tau,t}
\left(
d^t
\mid
y^{t-1},x^t
\right)
:=
Q_{\tau}
\left(
Y_{i,t}(d^t)
\,\middle|\,
Y_i^{t-1}(d^{t-1})=y^{t-1},
X_i^t=x^t
\right)
\]
be the $\tau$-quantile of the potential outcome under treatment history
$d^t$, conditional on the corresponding past potential-outcome
history and observed characteristics.

The following proposition links this object with the outcome function $\mu_{t}$. 

\begin{proposition}\label{prop:structural-quantile}
	Suppose Assumptions \ref{ass:DGP}--\ref{ass:U-DGP} hold.
	Then, for every $t\in\mathcal{T}$, every
	$d^t\in\{0,1\}^t$,
	\begin{equation}\label{eqn:structural-quantile}
		q_{\tau,t}
		\left(
		d^t
		\mid
		Y^{t-1}(d^{t-1}),X^t
		\right)
		=
		\mu_t
		\left(
		Y^{t-1}(d^{t-1}),
		X^t,
		d^t,
		\tau
		\right),~a.s.
	\end{equation}
	Moreover,
	\begin{equation}\label{eqn:observed-quantile-moment}
		E\left[
		1\left\{
		Y_{t}
		\leq
		\mu_t
		\left(
		Y^{t-1},
		X^t,
		D^t,
		\tau
		\right)
		\right\}
		-\tau
		\,\middle|\,
		Y^{t-1},
		X^t,
		D^t
		\right]
		=
		0,~a.s.
	\end{equation}
\end{proposition}

\begin{proof}
	See Appendix \ref{app:structural.QTE}.
\end{proof}

\paragraph{The null hypothesis.} We define the conditional structural quantile treatment effect of
profile $d^t$ relative to the control profile $0^t$ as
\begin{align}
	\operatorname{QTE}_{\tau,t}
	\left(
	d^t,
	y^{t-1},
	x^{t}
	\right)
	&:=
	q_{\tau,t}
	\left(
	d^t
	\mid
	y^{t-1},x^t
	\right)
	-
	q_{\tau,t}
	\left(
	0^t
	\mid
	y^{t-1},x^t
	\right) \notag \\
	&=
	\mu_t
	\left(
	y^{t-1},
	x^t,
	d^t,
	\tau
	\right)
	-
	\mu_t
	\left(
	y^{t-1},
	x^t,
	0^t,
	\tau
	\right).
	\label{eqn:conditional-qte}
\end{align}
This effect compares the two treatment profiles at the same history
of past outcomes and observed characteristics. If the outcome
function does not depend on past outcomes, it reduces to the usual
conditional quantile treatment effect given $X_i^t=x^t$.

Fix a set of treatment profiles
$\mathcal{D}^t\subseteq\{0,1\}^t\setminus\{0^t\}$ and a covariate
value $x^t$. Our null hypothesis is
\begin{equation}\label{eqn:null-hypothesis}
	H_0:
	\operatorname{QTE}_{\tau,t}
	\left(
	d^t,
	y^{t-1},
	x^t
	\right)
	=
	0,
\quad P_{d^t,x^t}\text{-a.e. }y^{t-1}, \quad \forall d^{t} \in \mathcal D^{t}.
\end{equation}
Equivalently,
\[
H_0:
\mu_t
\left(
\cdot,
x^t,
d^t,
\tau
\right)
=
\mu_t
\left(
\cdot,
x^t,
0^t,
\tau
\right),
\quad P_{d^t,x^t}\text{-a.e. }y^{t-1}, \quad \forall d^{t} \in \mathcal D^{t}.
\]


\subsection{Discussion of the Identifying Assumptions}
\label{sec:Discussion.Assumptions}

Assumption \ref{ass:DGP} is standard, imposing independence across units, but not
across time,  except that the law $P$ may depend on $n$. This allows the
probabilities of particular treatment profiles to approach zero as
the sample size increases; it accommodates settings in which some
treatment groups remain small even though the overall sample grows. The discrete covariates restriction can be somewhat relaxed; we discuss how in Section \ref{sec:reference-estimator}. The absolute continuity condition is technical; it ensures sufficient overlap to evaluate the reference-profile quantile function in the treated groups.  
Assumption \ref{ass:monotonicity} imposes
strict monotonicity of the outcome function in the latent rank, a
standard restriction in structural quantile models; e.g.
\citet{chernozhukov2005iv}.

The key identifying condition is Assumption \ref{ass:U-DGP}. To
describe its two distinct roles, define $H_t(d^t)
:=
\left(
Y^{t-1}(d^{t-1}),
X^t
\right)$. 
First, the within-profile restriction
\[
\Pr\left(
U_t(d^t)\leq\tau
\,\middle|\,
H_t(d^t),
D^t=d^t
\right)
=
\tau
\]
implies the observable conditional quantile moment for units whose
realized treatment history is $d^t$. This is the part of the
assumption needed for the validity of the testing procedure. Second, Assumption \ref{ass:U-DGP} requires
\[
\Pr\left(
U_t(d^t)\leq\tau
\,\middle|\,
H_t(d^t),
D^t=\widetilde d^t
\right)
=
\tau
\]
for every realized treatment profile $\widetilde d^t$, not only for $\widetilde d^t=d^t$. This cross-profile
restriction ensures that selection into treatment does not change the
conditional $\tau$-quantile of the latent rank associated with
$Y_t(d^t)$. It is this stronger restriction that permits the
conditional quantile identified from observable outcomes to be given
a causal interpretation.

A sufficient condition is $U_t(d^t)
\perp
D^t \mid
H_t(d^t)$, 
together with the normalization $\Pr\left(
U_t(d^t)\leq\tau
\,\middle|\,
H_t(d^t)
\right)
=
\tau$.

Randomized treatment assignment is an important special case, but it
is not the only case. Assumption \ref{ass:U-DGP} belongs to the broad class
of conditional-independence assumptions commonly referred to as
selection on observables or unconfoundedness (see, e.g., \citet{rosenbaum2010design}, \citet{imbens2000role}, and
\citet{firpo2007efficient}). Our condition is a quantile analogue of
these assumptions and, because it is imposed at a particular
$\tau$-quantile, is weaker than requiring the entire conditional
distribution of the potential rank to be invariant across treatment
profiles. Treatment assignment may depend arbitrarily on observed
covariates, past treatments, and past outcomes, provided that it is
independent of the relevant potential rank after conditioning on
these histories. The framework therefore covers randomized
experiments, observational designs satisfying selection on
observables, multivalued treatments, and dynamic assignments
satisfying an appropriate sequential conditional-independence
condition. For instance, Section \ref{sec:examples} below shows that it also accommodates
difference-in-differences designs under a quantile analogue of
parallel trends. When selection remains endogenous after conditioning on
the observed history, the assumption can also hold after including an
observed or estimable control variable, as in control-function
approaches to quantile regression; see \citet{lee2007endogeneity}.

Thus, Assumption \ref{ass:U-DGP} is substantive, but it is satisfied in many settings commonly encountered in applied research. Moreover, our model does not require treatment to be unconditionally exogenous, nor does it impose additive or homogeneous treatment effects, homoskedasticity, a parametric residual distribution, or a parametric specification of the outcome function. The framework therefore allows both treatment effects and residual distributions to vary flexibly across treatment histories and observed characteristics.

\subsection{Examples and Special Cases}
\label{sec:examples}

We now illustrate the scope of the framework and verify the identifying
assumptions in several familiar models. We begin with the
location-scale model from Section \ref{sec:illustrative}, then
consider a nonseparable quantile model and a
difference-in-differences design with staggered adoption.

\subsubsection{The Illustrative Location-Scale Model}
\label{sec:example-location-scale}

 For every treatment history
$d^t\in\{0,1\}^t$, suppose that
\begin{equation}\label{eqn:location-scale-potential}
	Y_t(d^t)
	=
	\theta_{d^t}
	+
	\gamma Y_{t-1}(d^{t-1})
	+ \lambda X^{t} + 
	\sigma_t\left(
	Y_{t-1}(d^{t-1}),
	X^t,
	d^t
	\right)
	\epsilon_t(d^t),
\end{equation}
where $\sigma_t(y_{t-1},x^t,d^t)>0$. Thus, the location and the scale of the potential outcome may vary with the lagged outcome, observed covariates, and treatment history --- importantly neither conditional homoskedasticity nor independence between the innovation and the lagged outcome is required.

For each $d^t$, let $F_{d^t}
\left(
e
\mid
y_{t-1},x^t
\right)
:=
P \left(
\epsilon_t(d^t)\leq e
\,\middle|\,
Y_{t-1}(d^{t-1})=y_{t-1},
X^t=x^t
\right)$ and let $U_t(d^t)
:=
F_{d^t}
\left(
\epsilon_t(d^t)
\,\middle|\,
Y_{t-1}(d^{t-1}),X^t
\right)$ be a latent rank variable. The corresponding outcome function is thus given by
\begin{align}
	\mu_t(y^{t-1},x^t,d^t,u)
	&:=
	\theta_{d^t}
	+
	\gamma y_{t-1} + \lambda x^{t} + 
	\sigma_t(y_{t-1},x^t,d^t)
	F_{d^t}^{-1}
	\left(
	u
	\mid
	y_{t-1},x^t
	\right),
	\label{eqn:location-scale-mu}
\end{align}

 We assume that $e \mapsto F_{d^t} \left( e \mid y_{t-1},x^t \right)$ is continuous and strictly increasing and impose the
normalization
\begin{equation}\label{eqn:error-quantile-normalization}
	F_{d^t}^{-1}
	\left(
	\tau
	\mid
	y_{t-1},x^t
	\right)
	=
	0
\end{equation}
for every $(y_{t-1},x^t,d^t)$. 

Because $\sigma_t(y_{t-1},x^t,d^t)>0$ and the conditional quantile
function is strictly increasing, Assumption \ref{ass:monotonicity} holds.

Assumption \ref{ass:U-DGP} requires that, for each potential treatment history $d^t$, conditional on past potential
outcomes and observed covariates, realized treatment assignment preserves
the $\tau$-quantile of $\epsilon_t(d^t)$ at zero. A sufficient condition to ensure this is
\begin{equation}\label{eqn:error-treatment-independence}
	\epsilon_t(d^t)
	\perp
	D^t \mid 
	Y_{t-1}(d^{t-1}),X^t.
\end{equation}


It follows from Proposition \ref{prop:structural-quantile} and
normalization \eqref{eqn:error-quantile-normalization} that
\begin{align*}
	q_{\tau,t}
	\left(
	d^t
	\mid
	y^{t-1},x^t
	\right)
	&=
	\theta_{d^t}
	+
	\gamma y_{t-1} + \lambda x^{t} + 
	\sigma_t(y_{t-1},x^t,d^t)
	F_{d^t}^{-1}
	\left(
	\tau
	\mid
	y_{t-1},x^t
	\right) =
	\theta_{d^t} + \lambda x^{t} 
	+
	\gamma y_{t-1}.
\end{align*}
Consequently,
\[
\operatorname{QTE}_{\tau,t}
\left(
d^t,y^{t-1}, x^t
\right)
=
\theta_{d^t}-\theta_{0^t}.
\]
Thus, for any collection $\mathcal{D}^t$ of treatment profiles, the
null hypothesis of no conditional quantile treatment effect is
equivalent to
\[
H_0:
\theta_{d^t}
=
\theta_{0^t}
\qquad
\text{for every }d^t\in\mathcal{D}^t.
\]

\paragraph{The linear regression case.}
The saturated specification permits the location parameter
$\theta_{d^t}$ to vary freely across treatment histories. A more
restrictive linear specification is
\[
\theta_{d^t}
=
\kappa+\beta'd^t,
\]
where $\beta=(\beta_{1},\ldots,\beta_{t})'$. Under this
specification, $H_{0} \colon \beta^{\prime} d^{t} = 0$ for all $d^{t} \in \mathcal D^{t}$.

\subsubsection{A Nonseparable Quantile Model}
\label{sec:example-nonseparable}

We next consider a nonseparable model in which treatment effects may
vary with observed characteristics, lagged outcomes, and latent
outcome ranks. This specification is related to the nonseparable structural quantile
models studied by \citet{chesher2003identification}, 
\citet{chernozhukov2005iv}, and to the nonseparable panel models of
\citet{chernozhukov2013average}. Its dynamic potential-outcome
structure is also related to \citet{han2018identification}. Our theory contributes to this literature by proposing inference when some
treatment-history groups contain only a small number of observations.

For every treatment history
$d^t\in\{0,1\}^t$, suppose that
\begin{align}
	Y_t(d^t)
	&=
	\kappa_t\left(
	X^t,
	U_t(d^t),
	Y_{t-1}(d^{t-1})
	\right) +	\theta_t\left(
	d^t,
	X^t,
	U_t(d^t),
	Y_{t-1}(d^{t-1})
	\right).
	\label{eqn:nonseparable-quantile}
\end{align}
The first component determines the potential outcome under the control
profile, while the second captures the effect of treatment history
$d^t$. We impose the normalization $\theta_t\left(
	0^t,x^t,u,y_{t-1}
	\right)
	=
	0$ for every $(x^t,u,y_{t-1})$.
	
The outcome function associated with
\eqref{eqn:nonseparable-quantile} is
\begin{align}
	\mu_t(y^{t-1},x^t,d^t,u)
	&=
	\kappa_t\left(
	x^t,u,y_{t-1}
	\right) +
	\theta_t\left(
	d^t,x^t,u,y_{t-1}
	\right).
	\label{eqn:nonseparable-mu}
\end{align}	

For each $d^t$, suppose that the latent rank satisfies
\begin{align}\label{eqn:nonseparable-rank-independence} 
	U_t(d^t)
	\perp
	D^t \mid
	Y_{t-1}(d^{t-1}),X^t,\quad and \quad 
	U_t(d^t) \mid
	Y_{t-1}(d^{t-1}),X^t
	\sim
	\mathcal{U}(0,1).
\end{align}
These conditions permit treatment assignment to depend on observed
characteristics and lagged outcomes. They require only that, after
conditioning on these variables, treatment assignment provide no
additional information about the latent rank of the potential
outcome.

We assume that, for every $(y_{t-1},x^t,d^t)$ in the relevant support, and any $d^{t} \ne 0^{t}$,
the mappings
\begin{equation}\label{eqn:nonseparable-monotonicity}
	u
	\mapsto
\kappa_t(x^t,u,y_{t-1})~and~u \mapsto 
	\theta_t(d^t,x^t,u,y_{t-1})
\end{equation}
are continuous and strictly increasing, so Assumption \ref{ass:monotonicity} holds. Moreover,
\eqref{eqn:nonseparable-rank-independence} implies that $\Pr\left(
U_t(d^t)\leq\tau
\,\middle|\,
Y_{t-1}(d^{t-1}),
X^t,
D^t
\right)
=
\tau$. 
Thus, Assumption \ref{ass:U-DGP} also holds.

It follows from Proposition \ref{prop:structural-quantile} that $q_{\tau,t}
	\left(
	d^t
	\mid
	y^{t-1},x^t
	\right)
	=
	\kappa_t\left(
	x^t,\tau,y_{t-1}
	\right) +
	\theta_t\left(
	d^t,x^t,\tau,y_{t-1}
	\right)$. Under our normalization, the conditional quantile
treatment effect is therefore $\operatorname{QTE}_{\tau,t}
	\left(
	d^t,y^{t-1},x^{t}
	\right)
	=
	\theta_t\left(
	d^t,x^t,\tau,y_{t-1}
	\right)$. For any collection of treatment profiles $\mathcal{D}^t$, the null
hypothesis of no conditional quantile treatment effect becomes
\[
H_0:
\theta_t\left(
d^t,x^t,\tau,y_{t-1}
\right)
=
0, \quad P_{d^{t},x^{t}}-a.e.~y^{t-1}, \quad d^t\in\mathcal{D}^t.
\]

\subsubsection{Difference-in-differences with staggered adoption}
\label{sec:example-DiD}

Recently, difference-in-differences designs have been applied to staggered treatment adoption settings (e.g. \cite{ChaiseMartin2020,SUN2020,callaway2018difference,athey2018designbased} and references therein). This framework is a particular case of ours where \(D_{i,t}\leq D_{i,t+1}\), so what matters is unit \(i\)'s treatment-adoption date $G_i=\min\{s:D_{i,s}=1\}$. We set \(G_i=0\) for
never-treated units and assume that no unit is treated in the initial
period. 

Let \(Y_{i,t}(g)\) denote the potential outcome at date \(t\) when
treatment is first received at date \(g\), and let \(Y_{i,t}(0)\) denote
the potential outcome under no treatment. Consider the fixed-effects
model
\begin{equation}
	Y_{i,t}(g)
	=
	\kappa_i+\delta_t
	+
	\sum_{\gamma=2}^{t}
	\theta_{t,\gamma}1\{g=\gamma\}
	+
	\zeta_{i,t},
	\qquad g\in\{0,2,\ldots,T\}.
	\label{eq:did-potential-outcomes}
\end{equation}
Here, \(\kappa_i\) is an individual fixed effect, \(\delta_t\) is a time
fixed effect, and \(\theta_{t,g}\) is the effect at date \(t\) of first
receiving treatment at date \(g\).

Using period \(0\) as the common pretreatment period, define $\Delta Y_{i,t}(g)
=
Y_{i,t}(g)-Y_{i,0}(g)$. 
Since no unit is treated at date \(0\), equation
\eqref{eq:did-potential-outcomes} gives
\begin{equation}
	\Delta Y_{i,t}(g)
	=
	c_t+\theta_{t,g}+\varepsilon_{i,t},
	\qquad t\geq g,
	\label{eq:did-differenced-model}
\end{equation}
where $c_t=\delta_t-\delta_0$, and $\varepsilon_{i,t}=\zeta_{i,t}-\zeta_{i,0}$.

We now embed \eqref{eq:did-differenced-model} in our framework and
verify the identifying assumptions.  For each cohort \(g\), let $F_{t,g}(e)
=
P\!\left(\varepsilon_{i,t}\leq e\mid G_i=g\right)$ 
denote the conditional distribution function of the change in the
idiosyncratic component. We assume that \(F_{t,g}\) is continuous and
strictly increasing and impose
\begin{equation}
	F_{t,g}^{-1}(\tau)
	=
	F_{t,0}^{-1}(\tau)
	=
	0
	\qquad
	\text{for every }g.
	\label{eq:did-quantile-parallel-trends}
\end{equation}
Condition \eqref{eq:did-quantile-parallel-trends} is a quantile analogue
of the standard parallel-trends condition. Indeed, whereas conventional
parallel trends requires $E\!\left[\varepsilon_{i,t}\mid G_i=g\right]
=
E\!\left[\varepsilon_{i,t}\mid G_i=0\right]$, 
condition \eqref{eq:did-quantile-parallel-trends} equates the
\(\tau\)-quantiles of the untreated outcome changes across cohorts.
The two conditions restrict different features of the counterfactual
distribution and neither generally implies the other.  Our condition also relates to assumptions
used to identify distributional treatment effects, including distributional
parallel trends combined with copula stability \cite{callawayli2019},
and monotone nonseparable models with time-invariant distributions of
unobservables within groups \cite{atheyimbens2006}.

For each
potential adoption date \(g\), define $U_{i,t}(g)
=
F_{t,g}(\varepsilon_{i,t})$
and
\begin{equation}
	\mu_t(g,u)
	=
	c_t+\theta_{t,g}+F_{t,g}^{-1}(u).
	\label{eq:did-structural-map}
\end{equation}
It follows immediately that $\Delta Y_{i,t}(g)
=
\mu_t\!\left(g,U_{i,t}(g)\right)$. 

Assumption \ref{ass:DGP} generalizes the standard assumption in this setup by allowing the sampling law to
depend on \(n\), so the probability of belonging to a particular
adoption cohort may approach zero. Assumption \ref{ass:monotonicity} follows because, for every \(g\), $u\mapsto F_{t,g}^{-1}(u)$ 
is continuous and strictly increasing.

To verify Assumption \ref{ass:U-DGP}, recall that by \eqref{eq:did-quantile-parallel-trends},
\(F_{t,g}^{-1}(\tau)=0\). Therefore, for any \(g\) and any realized
cohort \(\widetilde g\),
\begin{align*}
	&P\!\left(
	U_{i,t}(g)\leq\tau
	\mid G_i=\widetilde g
	\right) = 
	P\!\left(
	\varepsilon_{i,t}\leq F_{t,g}^{-1}(\tau)
	\mid G_i=\widetilde g
	\right) =
	P\!\left(
	\varepsilon_{i,t}\leq0
	\mid G_i=\widetilde g
	\right)
	=\tau.
\end{align*}
This is precisely the conditional rank
restriction in Assumption \ref{ass:U-DGP}. 

Evaluating \eqref{eq:did-structural-map} at \(u=\tau\) gives $\mu_t(g,\tau)
=
c_t+\theta_{t,g}$ and $\mu_t(0,\tau)
= c_t$. Consequently,
\begin{equation}
	\mu_t(g,\tau)-\mu_t(0,\tau)
	=
	\theta_{t,g} - \theta_{t,0}.
	\label{eq:did-structural-effect}
\end{equation}

As it turns out, in this model, our theory provides a testing procedure not only for QTE but also for ATE and ATT. To see this, observe that the treatment effect in \eqref{eq:did-potential-outcomes} is a constant
shift within each cohort-date pair. Hence, $\operatorname{ATE}(g,t)
=
E\!\left[Y_{i,t}(g)-Y_{i,t}(0)\right]
=
\theta_{t,g} - \theta_{t,0}$. 
Moreover, it is easy to show that the ATT and the QTT of group $g$ at date $t$ are also given by $\theta_{t,g}$. Thus, in this model the null hypotheses of no ATE/ATT, and no QTE all reduce to

\begin{equation}
	H_{0,t,g}:
	\theta_{t,g}= \theta_{t,0}. 
	\label{eq:did-null}
\end{equation}

\paragraph{Extensions to \eqref{eq:did-structural-map}.} The fixed-effects separable specification in \eqref{eq:did-structural-map} is used only to make the
connection with difference-in-differences transparent and is not required for quantile inference. One may allow the treatment effect to vary with the latent rank:
\[
\mu_t(g,u)
=
\mu_t(0,u)+\theta_{t,g}(u).
\]
The quantile treatment effect is then $\operatorname{QTT}_{\tau}(g,t)
=
\theta_{t,g}(\tau)$, and our framework directly tests $\theta_{t,g}(\tau)=\theta_{t,0}(\tau)$. In this more general model, however, treatment effects can differ across quantiles, and the ATE, ATT, and QTE need not coincide.

The framework also
allows the differenced potential outcome to depend on lagged outcomes
and time-varying covariates. For example, one may replace
\eqref{eq:did-differenced-model} with
\[
\Delta Y_{i,t}(g)
=
m_t\!\left(
Y_i^{t-1}(g),X_i^t
\right)
+
\theta_{t,g}
+
\varepsilon_{i,t}(g),
\]
where \(m_t\) may be nonlinear and the conditional distribution of
\(\varepsilon_{i,t}(g)\) may vary with the observed history. The
conditional quantile parallel-trends restriction is then imposed given
the relevant lagged outcomes and covariates, $H_{i,t}(g)
=
\left(Y_i^{t-1}(g),X_i^t\right)$. The same verification
applies conditionally, allowing for dynamic dependence, time-varying
covariates, heteroskedasticity, and richer treatment-history effects.

\section{Inference of Treatment Effects}
\label{sec:main}

In what follows, $t \in \mathcal{T}$ and $(x^{t},\tau) \in \mathcal{X}^{t} \cup \{\emptyset \} \times (0,1)$ and $\mathcal{D}^{t} \subseteq \{0,1\}^{t}$ are taken to be fixed. Also, we typically omit the $t$ index for the functions $\mu_{t}$ and $\delta_{t}$.\footnote{The symbol $\emptyset$ captures the case where there are no covariates.}

The goal is to construct a test for the aforementioned null hypothesis,
\begin{align*}
\mu(\cdot, x^{t},d^{t},\tau)  = \mu(\cdot, x^{t},0^{t},\tau), \quad  a.s.-P_{d^{t},x^{t}},~\forall d^{t} \in \mathcal{D}^{t}.
\end{align*}

The distinguishing feature of the problem is that the number of
observations associated with some profiles in \(\mathcal D^t\) need not
diverge. Because the number of treatment-profile groups is fixed, at
least one group must contain a nonvanishing fraction of the sample. We
assume throughout that this is the control group. That is, $\inf_n
	P_n\!\left(
	X^t=x^t,\,
	D^t=0^t
	\right)>0$. 
Consequently, the conditional quantile function associated with the
control profile can be consistently estimated, whereas the sizes of
the remaining treatment-profile groups may grow slowly or remain
bounded.\footnote{If a treatment-profile group, other than the control one, is asymptotically
	large, all our results extend to using this other group as the
	reference group.}

\paragraph{Overview.}
The test proceeds in three steps. First, we use the large reference
group \((x^t,0^t)\) to consistently estimate its structural conditional
quantile function, $\mu_t(\cdot,x^t,0^t,\tau)$. Second, under the null hypothesis, this function must also satisfy the
conditional quantile moments associated with every treatment profile
\(d^t\in\mathcal D^t\). We therefore evaluate the reference-group
estimate in the sample moments of those treatment-profile groups. Third,
we compare an aggregate of these moments with a critical value obtained
from their conditional Bernoulli distribution. This last distribution
is available without requiring the treatment-profile group sizes to
diverge. The oracle statistic has an exact conditional distribution,
while consistency of the reference-group estimator establishes the
asymptotic validity of the feasible test.

\paragraph{Notation.} In what follows, let $O^{t} : = (Y^{t-1},X^{t},D^{t})$. Also, it is useful to keep track of the units that belong to different treatment groups. A group is indexed by $g \in \mathcal{X}^{t} \cup \{ \emptyset \} \times \{0,1\}^{t} $, and let $$\mathcal{G}_{n}(g) : = \{ 1 \leq i \leq n \colon  (X^{t}_{i},D^{t}_{i}) = g  \}$$ be all units in a sample of $n$ units that belong to group $g$. As mentioned above, we ``normalize" the group $(x^{t},0^{t})$ to grow with the sample size $n$. Importantly, we allow for the size of other groups to grow slower than $n$ or not grow at all. Finally, let $M : = |\mathcal{D}^{t}|$ and sometimes we index the elements in $\mathcal{D}^{t}$ as $d^{t}[m]$ for $m \in \{1,...,M\}$, and for convenience $d^{t}[0] : = 0^{t}$.
Fix \(t\in\mathcal T\), \(x^t\in\mathcal X^t\), and
\(\tau\in(0,1)\), and let
\(\mathcal D_t\subseteq\{0,1\}^t\setminus\{0^t\}\) denote the treatment
profiles included in the null hypothesis. We suppress the \(t\) index
from \(\mu_t\) when no confusion can arise. 

For any $m \in \{0,...,M\}$,
\begin{align*}
	N_{n,m} : = N_n(g[m])=|\mathcal G_n(g[m])|~where~g[m] : = (x^{t},d^{t}[m]).
\end{align*}

Finally, we will sometimes use more shorthand notation: $\mu_{m}(y^{t-1}) : = \mu_{t}(y^{t-1},x^{t},d^{t}[m],\tau)$; $P_{m}(\cdot)$ denotes the conditional probability distribution of $Y^{t}$ given group $g[m]$; and $F_{m}(\cdot \mid y^{t-1})$ the conditional CDF of $Y^{t}$ given group $g[m]$ and $y^{t-1}$.

\subsection{Estimating the Reference-Profile Quantile Function}
\label{sec:reference-estimator}

Let \(\rho(y,q):=1\{y\leq q\}\). By
Proposition \ref{prop:structural-quantile}, the reference-profile structural quantile
function satisfies
\begin{align}
	E_P\!\left[
	\left\{
	\rho\!\left(
	Y_t,\mu(Y^{t-1},x^t,0^t,\tau)
	\right)-\tau
	\right\}
	\Phi_K(Y^{t-1})
	\,\middle|\,
	(X^t,D^t)=(x^t,0^t)
	\right]
	=0
	\label{eq:population-sieve-moment}
\end{align}
for every vector of integrable basis functions
\(\Phi_K:\mathcal Y^{t-1}\rightarrow\mathbb R^K\) and any $ K \in \mathbb N$.

For a candidate function \(h\), define the corresponding sample moment
\begin{align}
	\widehat{m}_{n,K}(x^t,0^t,h)
	:=
	\frac{1}{N_n(x^t,0^t)}
	\sum_{i\in\mathcal G_n(x^t,0^t)}
	\left\{
	\rho\!\left(Y_{i,t},h(Y_i^{t-1})\right)-\tau
	\right\}
	\Phi_K(Y_i^{t-1}).
	\label{eq:sample-reference-moment}
\end{align}
Let \(\mathcal H_K(0^t)\) be a sieve approximation to the function
class containing \(\mu(\cdot,x^t,0^t,\tau)\). We estimate the
reference-profile quantile function by\footnote{If the minimum is not attained, we use any measurable approximate
	minimizer.}
\begin{align}
	\widehat\mu_{n,K}(\cdot,x^t,0^t,\tau)
	\in
	\arg\min_{h\in\mathcal H_K(0^t)}
		\widehat{m}_{n,K}(x^t,0^t,h)'
	\widehat{m}_{n,K}(x^t,0^t,h).
	\label{eq:reference-profile-estimator}
\end{align}

Before moving on to the testing procedure, a few remarks are in order. First, the proposed estimator is a sieve minimum distance (see \cite{chen2009efficient,chen2012estimation}) and, because \(N_n(x^t,0^t)\) grows proportionally with \(n\) the standard asymptotic properties apply. However, any other estimator would also work provided it satisfies a consistency requirement presented in Assumption \ref{ass:oracle-approximation} below; indeed, in many cases a ``parametric" quantile regression is enough (see the examples below). Second, continuous covariates can be included by appending them alongside \(Y^{t-1}\), while continuing to define treatment-profile groups using the discrete covariates. Third, we impose no consistency requirement on quantile-function estimators constructed from the remaining treatment-profile groups, whose group sizes may grow slowly or remain bounded.

\subsection{The Testing Procedure}
\label{sec:testing-procedure}

%
%

Let \(\mathbf K=(K_0,\ldots,K_M)\). Our test statistic is constructed using a measure of distance between the treatment-profile
moments and zero. There are many possibilities; the leading case is given by
\begin{align}
	T_{n,\mathbf K}(h)
	:=
	\sum_{m=0}^M
	\left\|
	\widehat m_{n,K_m}(x^t,d^t[m],h)
	\right\|_2^2,
	\label{eq:sum-test-statistic}
\end{align}
but an alternative could be the maximum statistic\footnote{More generally, the results allow
	\[
	T_{n,\mathbf K}(h)
	=
	H\!\left(
	\widehat m_{n,K_0}(x^t,d^t[0],h),\ldots,
	\widehat m_{n,K_M}(x^t,d^t[M],h)
	\right),
	\]
	where \(H(0,\ldots,0)=0\) and, for some finite \(C\), $\left|
	H(a_0,\ldots,a_M)-H(b_0,\ldots,b_M)
	\right|
	\leq
	C\max_{0\leq m\leq M}
	\left|
	\|a_m\|_2^2-\|b_m\|_2^2
	\right|$. 
	This class includes the sum and maximum statistics, as well as
	weighted versions with uniformly bounded weights measurable with
	respect to \((O_i^t)_{i=1}^n\).}
\begin{align}
	T_{n,\mathbf K}(h)
	:=
	\max_{0\leq m\leq M}
	\left\|
	\widehat m_{n,K_m}(x^t,d^t[m],h)
	\right\|_2^2.
	\label{eq:max-test-statistic}
\end{align}
The sum statistic aggregates violations across treatment profiles,
whereas the maximum statistic is sensitive to the largest
profile-specific violation. The results below apply to either
choice.\footnote{The reference-profile moment is included in
	\(T_{n,\mathbf K}\) but just for completeness. The reference group is used to estimate \(\mu_0\);
	the null is tested by evaluating this function in the moments
	associated with the \emph{remaining} treatment profiles.}

Here and throughout, we assume that \(N_n(x^t,d^t[m])>0\) for all $m$ --- this assumption is made for simplicity; an empty treatment group would contain no information and would be omitted from the test statistic.

\paragraph{The oracle conditional distribution} We first consider the infeasible oracle statistic
\begin{align}
	T_{n,\mathbf K}^0
	:=
	T_{n,\mathbf K}(\mu(\cdot,x^t,0^t,\tau)),
	\label{eq:oracle-test-statistic}
\end{align}
which treats the reference-profile quantile function as known.

For each \(m\in\{0,\ldots,M\}\), define
\begin{align}
	\ell_{n,K_m}(d^t[m])
	:=
	\frac{1}{N_n(x^t,d^t[m])}
	\sum_{i\in\mathcal G_n(x^t,d^t[m])}
	(B_i-\tau)\Phi_{K_m}(Y_i^{t-1}),
	\label{eq:bernoulli-profile-moment}
\end{align}
where \(B_1,\ldots,B_n\) are IID
\(\mathrm{Bernoulli}(\tau)\) random variables independent of
\((O_i^t)_{i=1}^n\).

Let \(T_{n,\mathbf K}^B\) be defined by applying the same aggregation
rule used in \(T_{n,\mathbf K}\) to the Bernoulli moments in
\eqref{eq:bernoulli-profile-moment}. Thus, under the sum statistic,
\begin{align}
	T_{n,\mathbf K}^B
	=
	\sum_{m=0}^M
	\left\|
	\ell_{n,K_m}(d^t[m])
	\right\|_2^2,
	\label{eq:bernoulli-sum-statistic}
\end{align}
while under the maximum statistic,
\begin{align}
	T_{n,\mathbf K}^B
	=
	\max_{0\leq m\leq M}
	\left\|
	\ell_{n,K_m}(d^t[m])
	\right\|_2^2.
	\label{eq:bernoulli-max-statistic}
\end{align}

The following result gives the joint conditional distribution of the
treatment-profile moments.

\begin{lemma}
	\label{lem:conditional-bernoulli}
	Suppose Assumptions~\ref{ass:DGP}--\ref{ass:U-DGP} hold. Under \(H_0\), conditionally on
	\((O_i^t)_{i=1}^n\),
	\begin{align}
		\left(
		\widehat m_{n,K_m}(x^t,d^t[m],\mu_0)
		\right)_{m=0}^M
		\overset{d}{=}
		\left(
		\ell_{n,K_m}(d^t[m])
		\right)_{m=0}^M.
		\label{eq:joint-bernoulli-representation}
	\end{align}
	Consequently,
	\begin{align}
		T_{n,\mathbf K}^0 \mid 
		(O_i^t)_{i=1}^n
		\overset{d}{=}
		T_{n,\mathbf K}^B \mid 
		(O_i^t)_{i=1}^n.
		\label{eq:oracle-statistic-bernoulli-law}
	\end{align}
\end{lemma}

\begin{proof}
	Appendix \ref{app:conditional-bernoulli}. 
\end{proof}

The result holds for every \(n,\mathbf{K}\) and does not require the
treatment-profile group sizes to diverge. To see the underlying
argument, observe that, under \(H_0\), for every
\(i\in\mathcal G_n(x^t,d^t[m])\), $\rho\!\left(Y_{i,t},\mu_0(Y_i^{t-1})\right) 	=	1\{U_{i,t}(d^t[m])\leq\tau\}$.
Assumption~\ref{ass:U-DGP} implies that this indicator is conditionally
\(\mathrm{Bernoulli}(\tau)\), while Assumption~1 gives independence
across units. 

For \(\alpha\in(0,1)\), define the conditional critical value
\begin{align}
	c_{n,\mathbf K}(\alpha)
	:=
	\inf\left\{
	c:
	P\!\left(
	T_{n,\mathbf K}^B\leq c
	\,\middle|\,
	(O_i^t)_{i=1}^n
	\right)
	\geq 1-\alpha
	\right\}.
	\label{eq:bernoulli-critical-value}
\end{align}
In practice, this critical value can be approximated by repeatedly
drawing IID \(\mathrm{Bernoulli}(\tau)\) variables, constructing the
moments in \eqref{eq:bernoulli-profile-moment}, and computing the
corresponding value of \(T_{n,\mathbf K}^B\).

Lemma~\ref{lem:conditional-bernoulli} immediately implies
\begin{align}
	P\!\left(
	T_{n,\mathbf K}^0
	>
	c_{n,\mathbf K}(\alpha)
	\,\middle|\,
	(O_i^t)_{i=1}^n
	\right)
	\leq\alpha
	\qquad\text{a.s.}
	\label{eq:oracle-conditional-size}
\end{align}
Thus, if the reference-profile quantile function were known, the test
would control size conditionally in finite samples, including when
some treatment-profile groups contain only finitely many observations.

\paragraph{The Test Statistic} The oracle statistic is infeasible because \(\mu_0\) is unknown. Hence, 
we replace it with the reference-profile estimator constructed in
Section~\ref{sec:reference-estimator} and define
\begin{align}
	\widehat T_{n,\mathbf K(n)}
	:=
	T_{n,\mathbf K(n)}
	\left(
	\widehat\mu_{n,K_0(n)}(\cdot,x^t,0^t,\tau)
	\right),
	\label{eq:feasible-test-statistic}
\end{align}
where $T_{n,\mathbf{K}(n)}$ is as in \eqref{eq:max-test-statistic} or \eqref{eq:sum-test-statistic} with some sequence $\mathbf K(n)$ to be discussed below.

\subsubsection{Asymptotic Size} 

If replacing \(\mu_0\) with \(\widehat\mu_{n,K_0}\) is asymptotically negligible, then our previous results suggest the feasible test:
\begin{align}
	\varphi_{n,\mathbf K(n)}(\alpha)
	:=
	1\left\{
	\widehat T_{n,\mathbf K(n)}
	>
	c_{n,\mathbf K(n)}(\alpha)+\epsilon_n
	\right\}
	\label{eq:feasible-test}
\end{align}
should have correct asymptotic size. 

The quantity $\epsilon_{n}$ is presented below and accounts for the fact we do not know the oracle statistic.  We state this approximation requirement as a high-level condition (primitive sufficient conditions
are provided in Appendix \ref{app:oracle-sufficient-conditions} and Section \ref{sec:verification-assumption-4} below) and then state the formal asymptotic size result. 

\begin{assumption}
	\label{ass:oracle-approximation}
	There exist sequences $\mathbf K(n)=(K_0(n),\ldots,K_M(n))$, and	$\epsilon_n\downarrow0$ such that
	\begin{align}
		P\!\left(
		\left|
		\widehat T_{n,\mathbf K(n)}
		-
		T_{n,\mathbf K(n)}^0
		\right|
		>
		\epsilon_n
		\right)
		\longrightarrow0.
		\label{eq:oracle-approximation}
	\end{align}
\end{assumption}

Assumption~\ref{ass:oracle-approximation} restricts only the estimator
of the reference-profile function; it imposes no consistency
requirement on quantile-function estimators constructed from the
remaining treatment-profile groups. 

If $\mu$ has a parametric functional form, indexed by some finite-dimensional parameter, then under standard conditions $\epsilon_{n} = n^{-1/2} l_{n}$ for an arbitrarily slowly diverging sequence $(l_{n})_{n}$ such as $l_{n} = \log \log n$. However, if $\mu$ has a semi/nonparametric specification, $\epsilon_{n}$ will typically be slower than root-n reflecting (a) the standard nonparametric rates of estimation of $\mu$ for $d^{t}=0^{t}$ and (b) the potential growth of the tuning parameters $\mathbf{K}(n)$. See Appendix \ref{app:oracle-sufficient-conditions} for a more thorough discussion and references; see Section \ref{sec:verification-assumption-4} below for a low level conditions in the examples discussed in Section \ref{sec:examples}.


\begin{theorem}
	\label{thm:feasible-validity}
	Suppose Assumptions~\ref{ass:DGP}--\ref{ass:U-DGP} and
	Assumption~\ref{ass:oracle-approximation} hold. Then, under \(H_0\),
	for every \(\alpha\in(0,1)\),
	\begin{align}
		\limsup_{n\rightarrow\infty}
		E_P\!\left[
		\varphi_{n,\mathbf K(n)}(\alpha)
		\right]
		\leq\alpha.
		\label{eq:feasible-size-control}
	\end{align}
\end{theorem}

\begin{proof}
	See Appendix \ref{app:proof.size}. 
\end{proof}

To understand the result, consider the event $\left|
\widehat T_{n,\mathbf K(n)}
-
T_{n,\mathbf K(n)}^0
\right|
\leq\epsilon_n$. 
On this event, rejection by the feasible test implies $T_{n,\mathbf K(n)}^0
>
c_{n,\mathbf K(n)}(\alpha)$. 
Theorem~\ref{thm:feasible-validity} therefore follows from the
finite-sample conditional validity of the oracle test and the fact
that the probability of the complement of this event converges to
zero.

The result separates the two sources of uncertainty in the problem.
The Bernoulli critical value accounts exactly for the sampling
uncertainty in the treatment-profile groups, including groups whose
sizes remain bounded. Assumption~\ref{ass:oracle-approximation}
controls the estimation error introduced by replacing the unknown
reference-profile quantile function with
\(\widehat\mu_{n,K_0}\).

\subsection{Power}
\label{sec:power}

The following result formulates power relative to the information
contained in each treatment-profile group. The result applies without requiring every treatment-profile group to
grow with \(n\). 

For each \(l\in\{1,\ldots,M\}\), let \(\Delta_l\in\mathbb R_+\) denote
a lower bound on the treatment-profile-specific quantile effect:\footnote{The restriction to nonnegative effects is made only to simplify the
	presentation. An analogous result applies to uniformly nonpositive
	effects after reversing the relevant inequalities.}
\begin{align}
	\Delta_l
	\leq
	\Delta_l(Y^{t-1},X^t)
	:=
	\mu(Y^{t-1},X^t,d^t[l],\tau)
	-
	\mu(Y^{t-1},X^t,0^t,\tau),~a.s.
	\label{eqn:drift.ALT}
\end{align}

\begin{theorem}
	\label{thm:power}
	Suppose Assumptions \ref{ass:DGP}--\ref{ass:oracle-approximation} hold, and suppose that the first
	component of $\Phi_{K_l(n)}$ is equal to one and $a \mapsto F_{l}(a \mid y^{t-1},x^{t})$ is absolutely continuous
	with conditional density bounded below by $\underbar{f}_l > 0$  whenever $a$ lies between
	$\mu(y^{t-1},x^t,0^t,\tau)$ and
	$\mu(y^{t-1},x^t,d^t[l],\tau)$.

	Then, under expression \ref{eqn:drift.ALT},\footnote{Recall that $N_{n,m} = N_n(x^t,d^t[m])$ and $\mathcal G_{n,m} : = \mathcal G_{n}(x^{t},d^{t}[m])$ for any $m \in \{0,...,M\}$.}
	\begin{align*}
		 E\left[ 	\varphi_{n,\mathbf K(n)}(\alpha) \mid (O^{t})_{i=1}^{n}
		\right]  \geq & \max_{l \in \{1,...,M\}} \left\{  	\operatorname{F}_{\mathcal{N}(0,1)}\left(
		\frac{
			\sqrt{N_{n,l}} ( \underbar{f}_{l}  \Delta_{l} - \sqrt{c_{n}}) 
		}{
			\sigma_{n,l}
		}
		\right) -
		2C_{\mathrm{BE}}   \frac14
		\frac{
		 \left\|\Phi_{K_l(n)}\right\|^{3}_{L^{\infty}}
		}{
			\sqrt{N_{n,l}} \sigma_{n,l}^3
		} \right\}  - o_{P}(1), 
	\end{align*}
where $c_{n} : = c_{n,K(n)}(\alpha)+3\epsilon_{n}$ and 
\begin{align*}
  	\sigma_{n,l}^2 :=
  \frac{1}{N_{n,l}}
  \sum_{i\in\mathcal G_{n,l}}
  F_{l}(\mu_{0}(Y^{t-1}_{i})  \mid Y^{t-1}_i)(1-F_{l}(\mu_{0}(Y^{t-1}_{i})  \mid Y^{t-1}_i) )
  \left(  \frac{a^{\prime}_{n,l}}{||a_{n,l}||}  \Phi_{K_{l}}(Y^{t-1}_{i})   \right)^2 
\end{align*}
with $	a_{n,l}
:=
\frac{1}{N_{n,l}}
\sum_{i\in\mathcal G_{n,l}}
(F_{l}(\mu_{0}(Y^{t-1}_{i})  \mid Y^{t-1}_i)-\tau)  \Phi_{K_{l}}(Y^{t-1}_{i})  $.\footnote{In this case $\mu_0(y^{t-1}): = \mu(y^{t-1},x^{t},0^{t},\tau) $.}

%
\end{theorem}

\begin{proof}
	See Appendix \ref{app:power}.
\end{proof}

The additional conditions imposed in the theorem require, on the one hand, the first component of
\(\Phi_{K_l(n)}\) to be identically equal to one --- this requirement is essentially without
loss of generality, since the basis can always be augmented with a
constant. The lower bound on the conditional density is more
substantive. It could be somewhat relaxed, but is essentially necessary as it translates a separation in the conditional
quantile function into a separation in the conditional probability
that defines the tested moment. Without such a condition, a large
quantile treatment effect could occur over a region containing
arbitrarily little probability mass and therefore induce an
arbitrarily small displacement of the moment. In that case, no
uniform power guarantee stated directly in terms of \(\Delta_l\)
would be possible. 

The theorem shows how the power bound depends on the displacement of the
population moment relative to its sampling variability. To illustrate this, for each treatment
profile $l$, consider alternatives satisfying
\[
\underline f_l\Delta_l
=
\sqrt{c_n}
+
C\frac{\sigma_{n,l}}{\sqrt{N_{n,l}}},
\qquad C>0.
\]
If $\sigma_{n,l}$ is bounded away from zero, the lower bound simplifies to
\[
E\!\left[
\phi_{n,\mathbf K(n)}(\alpha)
\,\middle|\,
(O_i^t)_{i=1}^n
\right]
\ge
F_{\mathcal N(0,1)}(C)
-
O\!\left(
\frac{\|\Phi_{K_l(n)}\|_{L^\infty}^{3}}
{\sqrt{N_{n,l}}}
\right)
-o_P(1).
\]
Thus, when the remainder terms are small, the bound guarantees substantial
power if the effect-induced moment displacement exceeds the rejection
threshold by a sufficiently large multiple of its standard deviation.

These conclusions apply whether or not the treatment-profile group
grows with the overall sample size. When \(N_{n,l}\) is small or
bounded, the theorem provides a finite-sample lower bound, subject to
an explicit Berry--Esseen remainder measuring the accuracy of the
Gaussian approximation in that group. Unlike a conventional
local-power argument, the comparison therefore remains meaningful
when the amount of information in the group does not increase. In this
case, however, the sampling variability of the profile-specific
moment does not vanish, so alternatives detectable with substantial
power must remain separated from the null by a nonvanishing amount; the theorem provides a quantitative bound for this amount. 
When \(N_{n,l}\to\infty\), \(K_l(n)\) remains fixed, and the
directional variance is nondegenerate, the Berry--Esseen remainder
vanishes and the theorem recovers the familiar local-alternative
comparison. Moment departures of order \(N_{n,l}^{-1/2}\) generate
nondegenerate asymptotic power, whereas departures larger than
\(N_{n,l}^{-1/2}\) are detected with probability approaching one.

\subsection{Verification of Assumption \ref{ass:oracle-approximation} in the Examples}
\label{sec:verification-assumption-4}


\paragraph{The illustrative location-scale model.}
Consider the model in Section \ref{sec:example-location-scale}  whose structural outcome function is
\[
\mu_t(y^{t-1},x^t,d^t,u)
=
\theta_{d^t}
+
\lambda x^t
+ 
\gamma y_{t-1}
+
\sigma_t(y_{t-1},x^t,d^t)
F_{d^t}^{-1}(u\mid y_{t-1},x^t).
\]
The normalization $F_{d^t}^{-1}(\tau\mid y_{t-1},x^t)=0$ therefore implies that the reference-profile conditional quantile function is
\begin{align*}
	\mu_t(y^{t-1},x^t,0^t,\tau)
	=
	\theta_{0^t} + \lambda x^t +\gamma y_{t-1} = : \alpha_0(x^{t}) + \gamma y_{t-1}
\end{align*}
Thus, although the full conditional distribution is allowed to depend
nonparametrically on the history, the particular conditional quantile function
estimated from the reference group is finite-dimensional, indexed by $(\alpha_{0}(x^{t}), \gamma )$. Hence, one can hold $\mathbf{K}(n)$ fixed as $n$ grows and employ standard quantile regression methods (e.g. \cite{koenker1978regression}) to estimate these parameters at root-n rate.

\begin{proposition}
	\label{prop:oracle-location-scale}
	Suppose the conditions in
	Section~\ref{sec:example-location-scale} hold. Suppose further that for every \(m\in\{0,\ldots,M\}\),
	
	\begin{enumerate}
		\item $	\sigma_t(y_{t-1},x^t,d^t[m]) \geq c$ for some universal $c>0$.
		
		\item The conditional CDF of
		\(\epsilon_t(d^t[m])\) given \((Y^{t-1},X^t)\) is Lipschitz in its
		first argument with a common finite constant \(A\).
		
	\end{enumerate}
	
	Then Assumption~\ref{ass:oracle-approximation} holds with $\epsilon_n=(\log\log n)n^{-1/2}$. 
\end{proposition}

\begin{proof}
	See Appendix \ref{app:Suff.Oracle.Examples}.
\end{proof}

\paragraph{Difference-in-differences with staggered adoption} For the model in Section \ref{sec:example-DiD} the results are analogous.  For cohort \(g\), the differenced outcome satisfies $\Delta Y_{i,t}(g)=c_t+\theta_{t,g}+\epsilon_{i,t}$. Mathematically, this model is a particular case of the location-scale model but using $\Delta Y_{i,t}$ as the outcome variable instead of $Y_{i,t}$. Thus, the outcome function for the control group can be estimated at root-n rate by standard methods. Applying Proposition \ref{prop:oracle-location-scale} to this case shows Assumption \ref{ass:oracle-approximation} holds with $\epsilon_{n} = (\log \log n) n^{-1/2}$.

\paragraph{Non-separable quantile model} The model is fully described by  $	g_m(u,y^{t-1})
:=
\mu_{t}(y^{t-1},x^{t},d^{t}[0],u) 
+
\theta_t(d^t[m],x^t,u,y^{t-1})$. We assume that $y^{t-1} \mapsto \mu_{t}(y^{t-1},x^{t},d^{t}[0],\tau) $ belongs to a $s$-smooth class such as a H\"{o}lder class. Under this assumption (and other technical conditions common in the literature) standard sieve estimation techniques with dimension $J_{n} = n^{d/(2s+d)}$ (see e.g. \cite{chen2012estimation,chen2009efficient}) yield
	\[
\|\widehat\mu_{n,K_0(n)}-\mu_0\|_{L^2(P_{Y^{t-1}})}
=
O_P(r_n),
\qquad
r_n:=n^{-s/(2s+d)}.
\]

The next proposition leverages this result with other technical conditions to show that $\epsilon_{n} = (\log \log n) r_{n}$ when treatment classes, other than the control one, are finite. Henceforth, let $Q(\cdot \mid x^{t},d^{t}[m]) : = P(Y^{t-1} \in \cdot \mid (X^{t},D^{t})=(x^{t} , d^{t}[m]) )$.

\begin{proposition}
	\label{prop:oracle-nonseparable}
	Suppose the conditions in
	Section \ref{sec:example-nonseparable} and above hold. Suppose further that
	
	\begin{enumerate}
		\item There exists \(c_\mu>0\) such that, uniformly over
		\(m\), \(y^{t-1}\), and \(0<u_1<u_2<1\),
		\[
		g_m(u_2,y^{t-1})-g_m(u_1,y^{t-1})
		\geq c_\mu(u_2-u_1).
		\]

		\item The profile distributions are uniformly dominated by the
		marginal distribution: for some \(C<\infty\), $\frac{d Q(\cdot \mid x^{t},d^{t}[m])  }{dP_{Y^{t-1}} }\leq C$, for $m=0,\ldots,M$ and \(P_{X^t}\)-almost every \(x^t\).
		
		\item The dimensions \(K_m\), \(m\geq1\), are fixed and $\max_{1\leq m\leq M}
		\left\|
		\|\Phi_{K_m}\|_2
		\right\|_{L^2(P_{Y^{t-1}})}
		<\infty$. 
	\end{enumerate}
	
	If the reference-profile moment is omitted from the test statistic, then
	Assumption~\ref{ass:oracle-approximation} holds with $\epsilon_n=(\log\log n)n^{-s/(2s+d)}$. 
\end{proposition}

\begin{proof}
	See Appendix \ref{app:Suff.Oracle.Examples}.
\end{proof}

\subsection{Confidence Regions for Parametrized Treatment Effects}
\label{sec:confidence-regions}

We conclude the theoretical part by pointing out that the preceding testing procedure can also be inverted to construct
confidence regions for finite-dimensional parameters indexing the
structural quantile treatment effects. Let
\(\theta\in\Theta\subseteq\mathbb R^p\), and suppose that
\begin{align}
	\mu_t\!\left(y^{t-1},x^t,d^t[m],\tau\right)
	=
	\mu_t\!\left(y^{t-1},x^t,0^t,\tau\right)
	+
	g_m\!\left(y^{t-1},x^{t};\theta\right),
	\quad m\in\{1,\ldots,M\},
	\label{eq:indexed-treatment-effects}
\end{align}
where the functions \(g_m(\cdot;\theta)\) are known up to the
finite-dimensional parameter \(\theta\). Thus, $g_m\!\left(y^{t-1},x^{t};\theta\right)
=
\operatorname{QTE}_{\tau,t}
\!\left(d^t[m],y^{t-1},x^{t} \right)$, 
whereas the reference-profile quantile function
\(\mu_t(\cdot,x^t,0^t,\tau)\) is treated as an unrestricted nuisance
function and estimated from the reference-profile group.

The null hypothesis of no structural quantile treatment effects is
thus given by
\(g_m(\cdot;\theta)=0\) for every \(m\).

Let \(\widehat T_{n,\mathbf K}(\theta)\) be obtained by applying the
same aggregation rule used to construct \(T_{n,\mathbf K}\) to the
collection $\left(
\widehat m_{n,K_m}(d^t[m];\theta)
\right)_{m=1}^M$,
where, for each \(m\in\{1,\ldots,M\}\), 
\begin{align}
	\widehat m_{n,K_m}\!\left(d^t[m];\theta\right)
	:=
	\frac{
		\sum_{i\in\mathcal G_n(x^t,d^t[m])}
		\left[
		\rho\!\left(
		Y_{i,t},
		\widehat\mu_{n,K_0}
		(Y_i^{t-1},x^t,0^t,\tau)
		+
		g_m(Y_i^{t-1},x^{t};\theta)
		\right)
		-\tau
		\right]
		\Phi_{K_m}(Y_i^{t-1})}{N_n(x^t,d^t[m])}.
	\label{eq:indexed-sample-moment}
\end{align}

The \(1-\alpha\) confidence region is obtained as follows:
\begin{align}
	\operatorname{CR}_n(1-\alpha)
	:=
	\left\{
	\theta\in\Theta:
	\widehat T_{n,\mathbf K(n)}(\theta)
	\leq
	c_{n,\mathbf K(n)}(\alpha)+\epsilon_n
	\right\}.
	\label{eq:confidence-region}
\end{align}

Let \(\theta_0\) denote the true value, so that
\eqref{eq:indexed-treatment-effects} holds with \(\theta=\theta_0\).
Validity of the test at \(\theta_0\) immediately yields the following
coverage result.

\begin{corollary}[Coverage of the confidence region]
	\label{cor:confidence-region}
	Suppose that the conditions of
	Theorem~\ref{thm:feasible-validity} hold. Then
	\begin{align}
		\liminf_{n\rightarrow\infty}
		P\!\left(
		\theta_0\in\operatorname{CR}_n(1-\alpha)
		\right)
		\geq 1-\alpha.
		\label{eq:confidence-region-coverage}
	\end{align}
\end{corollary}

This corollary can readily be used to compute confidence intervals in examples in sections \ref{sec:example-location-scale} and \ref{sec:example-DiD} above.

\section{Monte Carlo Simulations}
\label{sec:MC}

We use a Monte Carlo exercise to evaluate the finite sample performance of our testing procedure, in particular its size and power. Given the nature of our setup, we put special emphasis on evaluating this for different distributions of the residuals. We deliberately use a simple location model so that
these mechanisms can be isolated transparently. Consider
\begin{align}
	Y_i
	=
	\theta_0+\theta_1D_i+U_i,
	\label{eq:mc-model}
\end{align}
where \(D_i=1\) identifies the treatment group. For each sample size
\(n\), we draw $D_i\sim\operatorname{Ber}(p_n)$ and $p_n=\frac{N_1}{n}$, so that the expected number of treated observations is \(N_1\). We
consider \(N_1\in\{5,10,20\}\). Thus, as \(n\) increases while \(N_1\)
remains fixed, the size of the reference group grows, whereas the
expected size of the treatment group does not. This reproduces the
asymptotic configuration studied in the paper.\footnote{Related Monte
	Carlo evidence on the performance of standard procedures when the
	number of treated observations is small is provided by
\cite{ferman2019inference}.}

To assess robustness to the distribution of the outcome disturbance,
we consider the following four specifications:
\begin{align*}
	\text{(i) Gaussian:}\qquad&
	U_i\sim N(0,1);
	\\
	\text{(ii) Gaussian random scale:}\qquad&
	U_i\mid Z_i
	\sim
	N\left(
	0,
	1+\sqrt{Z_i}+\left|\sin(\pi Z_i)\right|
	\right),
	\qquad
	Z_i\sim U(0,1);
	\\
	\text{(iii) Uniform:}\qquad&
	U_i\sim U(-\sqrt{3},\sqrt{3});
	\\
	\text{(iv) Fat tails:}\qquad&
	U_i\sim U(-v,v)
	\quad\text{with probability }1-\varepsilon,
\end{align*}
while, with probability \(\varepsilon\), \(U_i\) takes the values
\(-A\) and \(A\) with equal probability, where $v=0.1$, $\varepsilon=0.04$, and $A
=
\sqrt{
	\frac{1-(1-\varepsilon)(2v)^2/12}{\varepsilon}
}
\approx 5$. 
The last specification combines a large concentration of probability
near zero with a small probability of extreme realizations. In all
four designs, the median of \(U_i\) is zero. We set
\(\theta_0=\theta_1=0\), focus on the median
\(\tau=0.5\), and use \(10{,}000\) Monte Carlo repetitions. 

\paragraph{Size} The null hypothesis is $\theta_1 = 0$. We use \(d[0]=0\) as the reference profile and \(d[1]=1\) as the
treatment profile. We estimate the reference-profile median
\(\theta_0\) using only observations in \(\mathcal G_n(0)\), and denote the
resulting estimator by \(\widehat\theta_0\).

The test statistic used in the simulations is
\begin{align}
	\widehat Q_n
	&=
	\left[
	\frac{1}{N_n(0)}
	\sum_{i\in \mathcal G_n(0)}
	\left\{
	1\{Y_i\leq\widehat\theta_0\}-\frac12
	\right\}
	+
	\frac{1}{N_n(1)}
	\sum_{i\in \mathcal G_n(1)}
	\left\{
	1\{Y_i\leq\widehat\theta_0\}-\frac12
	\right\}
	\right]^2.
	\label{eq:mc-feasible-statistic}
\end{align}

To construct the oracle reference distribution, let
\((B_i)_{i=1}^n\) be IID Bernoulli random variables with success
probability \(1/2\), drawn conditionally on
\((D_i)_{i=1}^n\), and define $\ell_n(d)
:=
\frac{1}{N_n(d)}
\sum_{i\in \mathcal G_n(d)}
\left(B_i-\frac12\right)$, for $d\in\{0,1\}$. 
The conditional critical value is $c_n(\alpha)
:=
\inf
\left\{
c\geq0:
\Pr\left(
(\ell_{n}(0)  + \ell_{n}(1) )^{2} \leq c
\mid (D_i)_{i=1}^n
\right)
\geq1-\alpha
\right\}$. 
In the simulations, we approximate this conditional distribution
using 500 Bernoulli draws.

Table \ref{tab:mc_size} reports rejection probabilities for
\(n\in\{25,50,200,800\}\), expected treatment-group sizes
\(N_1\in\{5,10,20\}\), and nominal levels of \(1\%\), \(5\%\), and
\(10\%\). 

For \(n=200\) and \(n=800\), rejection rates are generally close to
their nominal levels, including when the treatment group contains
only five or ten observations on average. Thus, it is enough for the
reference group to be moderately large for the estimation of
\(\theta_0\) to have a negligible effect on the test statistic. This is
precisely the oracle-approximation mechanism established by
Theorem \ref{thm:feasible-validity} with $\epsilon_{n} = n^{-1/2} l_{n}$ for some slowly diverging $l_{n}$, e.g., $l_{n} = \log n$. 

A second notable feature is the stability of the rejection rates
across the four disturbance distributions. In particular, the test
behaves similarly under Gaussian, uniform, random-scale, and
fat-tailed disturbances. This reflects the distribution-free nature
of the oracle quantile indicators: under the null, they are
Bernoulli\((1/2)\), regardless of the remaining shape of the
disturbance distribution.

The procedure becomes conservative when the reference group is
itself small. This occurs for \(n=25\) and for \(n=50\) and
\(N_1=20\), where neither group provides a large reference sample.
Rejection probabilities in these designs are generally two to three
percentage points below their nominal levels. These cases clarify an
important distinction. The procedure permits the treatment group to
remain small, but oracle approximation still requires the
reference-profile estimator to be sufficiently accurate. The
conservativeness observed when both groups are small is consistent
with this requirement. For \(n=25\), we do not report the case
\(N_1=20\), because the expected size of the reference group is then
smaller than the expected size of the treatment group.


\begin{table}[!htb]
	\caption{Nominal Size and Simulated Rejection Rates}
	\begin{adjustwidth}{-1.5cm}{}
		{\footnotesize{\begin{tabular}{c|c|cccc|cccc|cccc}
					\hline \hline
					& & \multicolumn{4}{c}{\multirow{1}{*}{ $N_{1} = 5 $  }}& \multicolumn{4}{c}{\multirow{1}{*}{ $N_{1} = 10 $  }}  & \multicolumn{4}{c}{\multirow{1}{*}{ $N_{1} = 20 $  }}        \\ \hline 
					& Size/Model & (i) & (ii) & (iii) & (iv)   & (i) & (ii) & (iii) & (iv)  & (i) & (ii) & (iii) & (iv)         \\ \hline 
					\multirow{3}{*}{$n=25$}& $0.01$ & $0.002$ & $0.002$ & $0.002$ & $0.002$ &  $0.002$ & $0.003$ & $0.001$ & $0.003$   &  &  &  &   \\
					& $0.05$ & $0.03$ & $0.03$ & $0.03$ & $0.03$ &  $0.03$ & $0.03$ & $0.03$ & $0.03$   & &  & &    \\ 
					& $0.10$ & $0.07$ & $0.06$ & $0.06$ & $0.07$ &  $0.07$ & $0.06$ & $0.07$ & $0.07$    &  &  &  &   \\ \hline 
					\multirow{3}{*}{$n=50$}& $0.01$ & $0.005$ & $0.01$ & $0.004$ & $0.005$ & $0.01$ & $0.01$ & $0.01$ & $0.01$ & $0.005$ & $0.01$ & $0.004$ & $0.01$  \\
					& $0.05$ & $0.04$ & $0.04$ & $0.04$ & $0.04$ & $0.04$ & $0.04$ & $0.04$ & $0.04$ & $0.04$ & $0.03$ & $0.03$ & $0.03$  \\ 
					& $0.10$ & $0.08$ & $0.08$ & $0.08$ & $0.08$ & $0.08$ & $0.08$ & $0.08$ & $0.08$ & $0.07$ & $0.07$ & $0.07$ & $0.07$  \\ \hline
					\multirow{3}{*}{$n=200$}&$0.01$ & $0.01$ & $0.01$ & $0.01$ & $0.01$  & $0.01$ & $0.01$ & $0.01$ & $0.02$ & $0.01$ & $0.01$ & $0.01$ & $0.01$   \\ 
					&   $0.05$ & $0.04$ & $0.04$ & $0.04$ & $0.04$  & $0.05$ & $0.05$ & $0.04$ & $0.05$ & $0.05$ & $0.05$ & $0.05$ & $0.05$   \\ 
					&  $0.10$ &  $0.09$ & $0.08$ & $0.08$ & $0.08$ & $0.10$ & $0.09$ & $0.09$ & $0.09$ & $0.09$ & $0.09$ & $0.09$ & $0.09$ \\ \hline 
					\multirow{3}{*}{$n=800$} &$0.01$ & $0.01$ & $0.01$ & $0.01$ & $0.01$ & $0.01$ & $0.01$ & $0.01$ & $0.01$ & $0.01$ & $0.01$ & $0.01$ & $0.01$  \\
					&$0.05$ & $0.04$ & $0.04$ & $0.04$ & $0.04$ & $0.04$ & $0.05$ & $0.04$ & $0.05$ & $0.05$ & $0.05$ & $0.05$ & $0.05$ \\ 
					& $0.10$ & $0.08$ & $0.07$ & $0.08$ & $0.08$ & $0.09$ & $0.09$ & $0.09$ & $0.09$ & $0.10$ & $0.10$ & $0.10$ & $0.10$  \\ \hline                                         	\hline
		\end{tabular} }}
	\end{adjustwidth}
	\label{tab:mc_size}
	\footnotetext{Notes: This table reports the rejection rates of our Monte Carlo simulations for different test sizes, sample sizes $n$, expected number of treated units $N_{1}$,  and specification of the error term in the model: (i) $U \sim N(0,1)$; (ii) $U \sim N(0, 1+\sqrt{Z}+ | \sin(\pi Z)| )$ where $Z \sim U(0,1)$; (iii) $U \sim U(-\sqrt{3},\sqrt{3})$ and (iv) with probability $1-\epsilon$, $U \sim U(-v,v)$ and with probability $\epsilon$, $U  \in \{  -A, A \}$  with probability 0.5  where $v : = 0.1$, $\epsilon : = 0.04$ and $A: = \sqrt{ (1 - (1-\epsilon) (2 v)^2/12)/\epsilon} \approx 5$.}
\end{table}

The main conclusion from Table \ref{tab:mc_size} is that 
holding the expected treatment-group size fixed, the feasible test
approaches its oracle behavior as the reference group grows. The
simulations also confirm that this approximation is largely
insensitive to the shape and tails of the disturbance distribution.

\paragraph{Power and the Information Content of Small Treatment Cells}

\begin{figure}[h!]
	\caption{Simulated power functions}
	\begin{minipage}{.5\textwidth}
		\centering
		\subcaption*{i) Normal}
		\includegraphics[width=\linewidth,height=3.5cm]{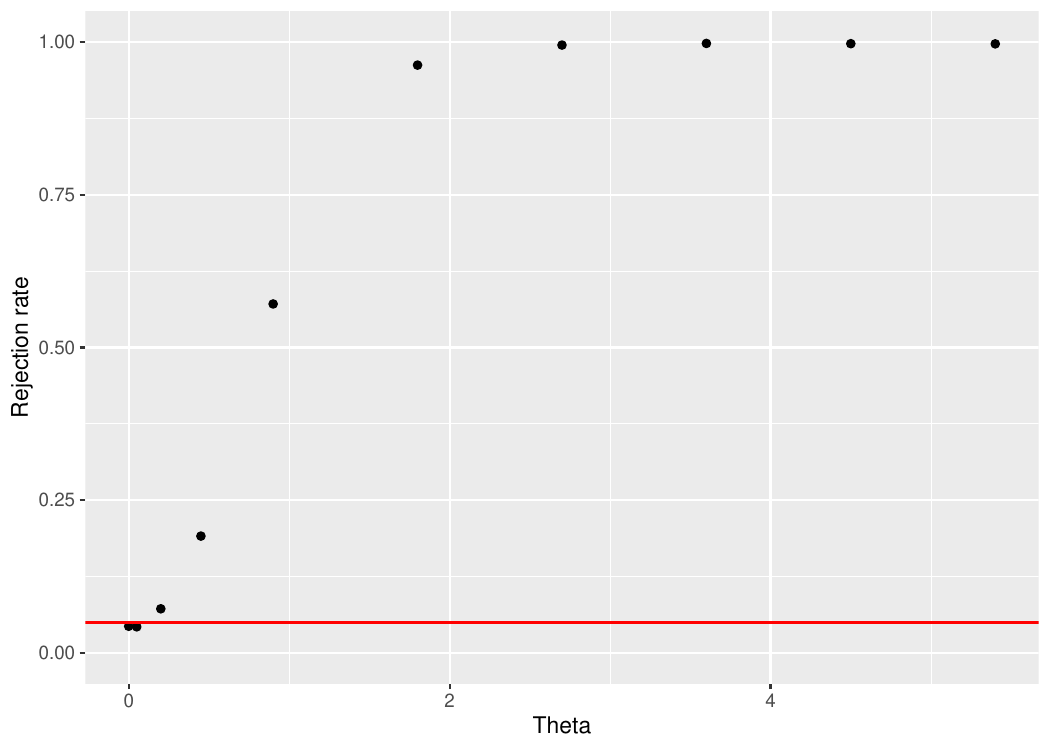}
	\end{minipage}%
	\begin{minipage}{.5\textwidth}
		\centering
		\subcaption*{ii) Heteroskedasticity}
		\includegraphics[width=\linewidth,height=3.5cm]{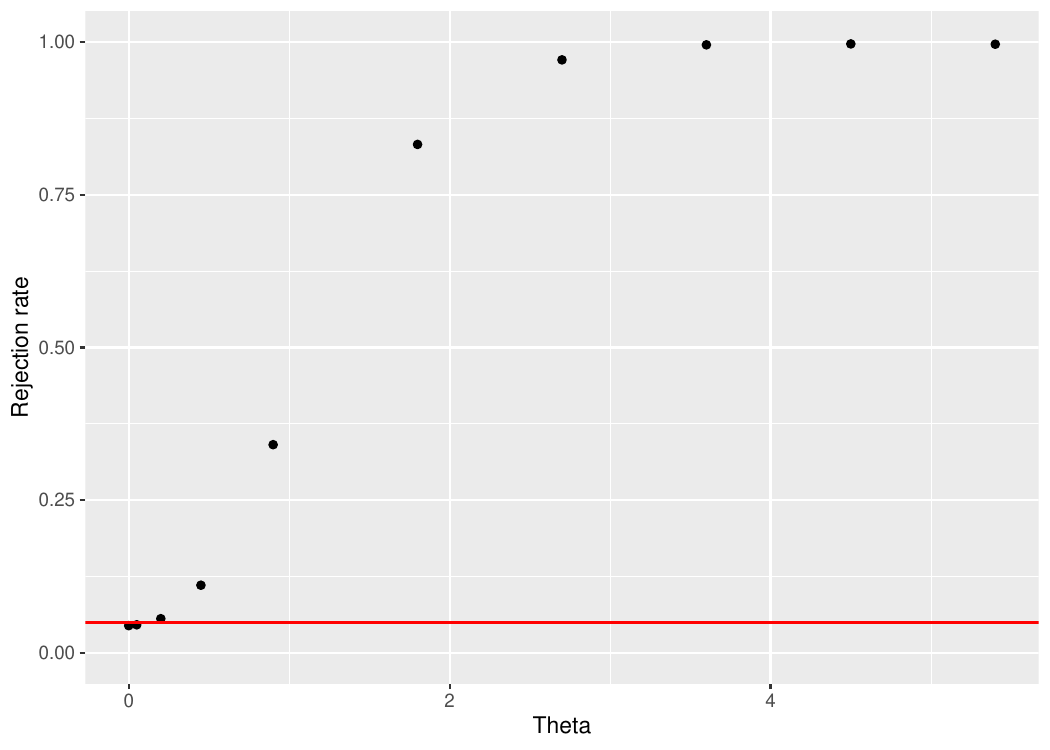}
	\end{minipage}%
	\\
	\begin{minipage}{.5\textwidth}
		\centering
		\subcaption*{iii) Uniform}
		\includegraphics[width=\linewidth,height=3.5cm]{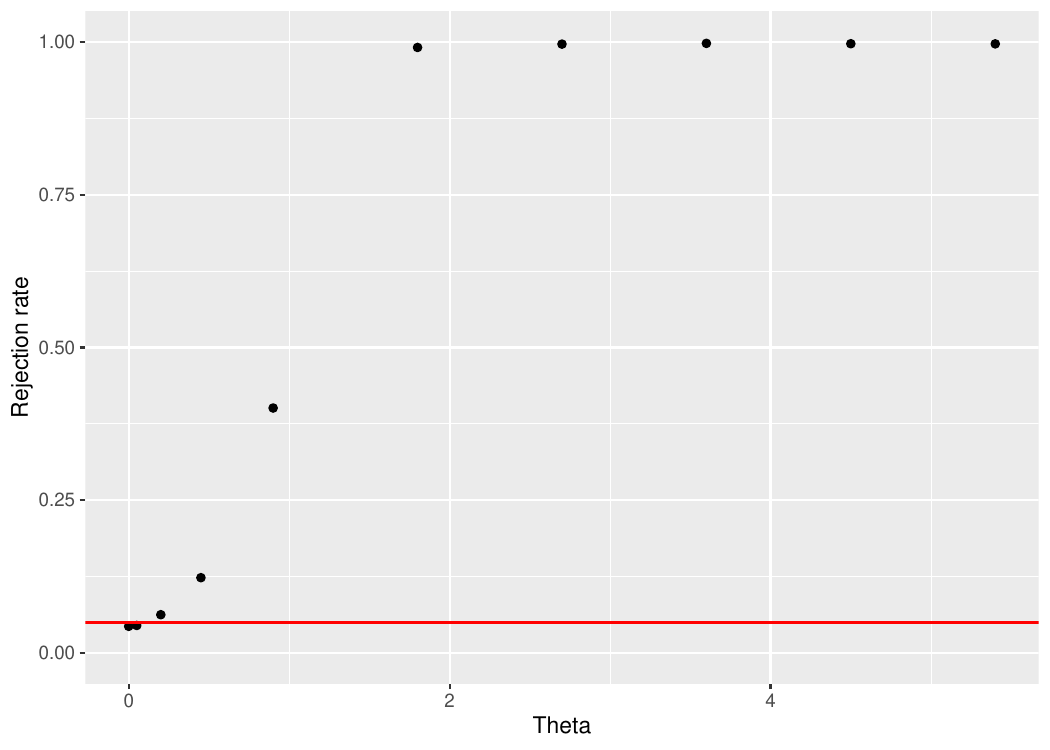}
	\end{minipage}%
	\begin{minipage}{.5\textwidth}
		\centering
		\subcaption*{iv) Fat tails}
		\includegraphics[width=\linewidth,height=3.5cm]{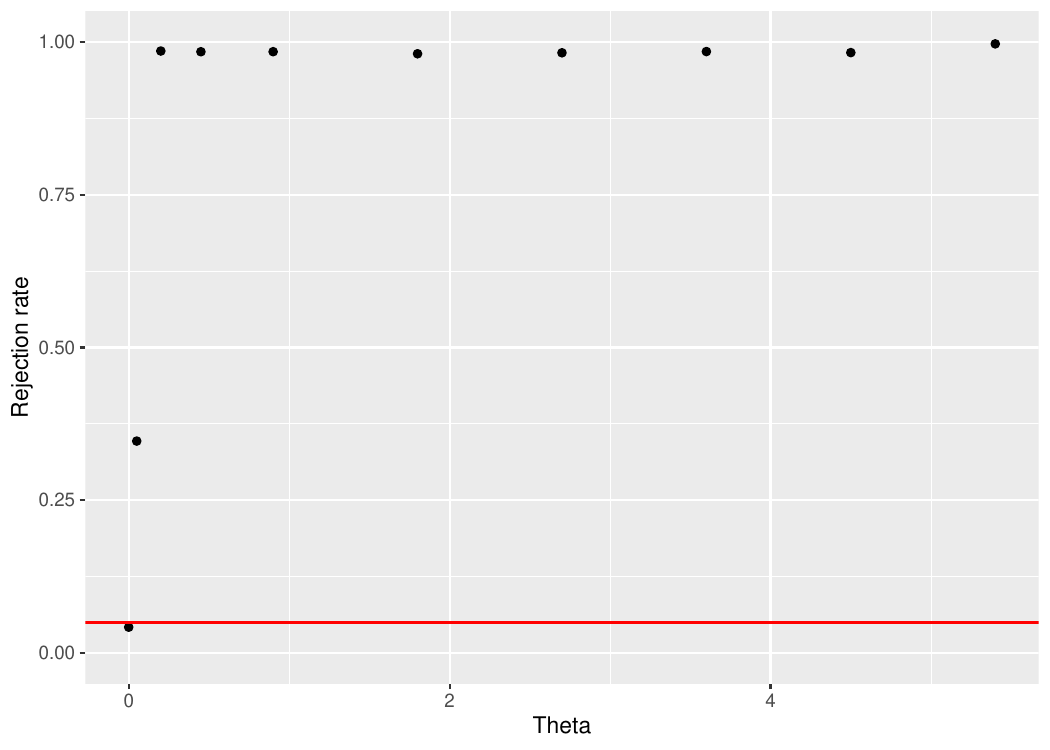}
	\end{minipage}%
	\footnotetext{Notes: In this figure we show the power functions that we estimate in our Monte Carlo exercise with nominal size $\alpha=0.05$. We choose $\pi = 0.95$ and $n = 200$, so the treatment group has, on average, 10 observations. }
	\label{fig:power}	
\end{figure}

Figure \ref{fig:power} reports simulated power at the \(5\%\)
significance level for \(n=200\) and \(N_1=10\). The reference group
therefore contains approximately 190 observations, whereas the
treatment group contains approximately ten. We vary $\theta_{1} : = \Delta  \in [0,5.4]$.

Power increases with the magnitude of the alternative and approaches
one in each design as the alternative becomes sufficiently large.
The rate at which power increases, however, differs across
disturbance distributions. To understand this pattern, consider the
treated-profile moment under the alternative:
\begin{align}
	\widehat m_n(1,\widehat\theta_0)
	&=
	\frac{1}{N_n(1)}
	\sum_{i\in \mathcal G_n(1)}
	\left[
	1\left\{
	U_i
	\leq
	\widehat\theta_0-\theta_0-\Delta
	\right\}
	-\frac12
	\right].
	\label{eq:mc-treated-moment}
\end{align}

Ignoring the estimation error in $\widehat\theta_0$, let $p(\Delta) :=
	F_U(-\Delta)$,  $\kappa(\Delta) :=
	p(\Delta)-\frac{1}{2}$, and $\sigma^2(\Delta)
	:=
	p(\Delta)\{1-p(\Delta)\}$. The power bound in
Section~\ref{sec:power} depends on the standardized distance between
the moment signal and the critical threshold, $\frac{\sqrt{N_{n}(1)}
		\left\{|\kappa(\Delta)|-\sqrt{c_{N_{n}(1)}(\alpha)}\right\}}
	{\sigma(\Delta)}$, 
together with the corresponding Berry--Esseen remainder. Power
therefore becomes substantial when the moment displacement exceeds
the critical threshold by an amount that is large relative to the
finite-sample variability of the treated-profile moment. It is
not $\Delta$ alone, or even $\sqrt{N_{n}(1)}|\kappa(\Delta)|$ alone, that
determines the rejection probability.

The four disturbance distributions translate a given
location shift $\Delta$ into different values of
$|\kappa(\Delta)|$, $\sigma^2(\Delta)$, and the Berry--Esseen
remainder. In particular, the fat-tailed specification places most of
its probability mass in a narrow interval around zero. A moderate location shift
therefore produces a comparatively large displacement of the median
moment and a rapid increase in power. Tail thickness by itself is not
the relevant determinant; what matters is how the distribution
around the tested quantile determines the signal relative to its
finite-group variability.

The power exercise also illustrates an unavoidable limitation of
inference with extremely sparse treatment groups. To see this,
consider a simplified test based only on the treatment-profile
moment. Conditional on \(N_n(1)\), its oracle counterpart is $\mathcal B_\tau(N_n(1))
:=
\left|
\frac{\operatorname{Bin}(N_n(1),\tau)}{N_n(1)}-\tau
\right|$, 
where \(\operatorname{Bin}(N_n(1),\tau)\) denotes a binomial random
variable with \(N_n(1)\) trials and success probability \(\tau\). Let $c_\alpha(N_n(1))
:=
\inf
\left\{
c\geq0:
\Pr\bigl(\mathcal B_\tau(N_n(1))\leq c\bigr)
\geq1-\alpha
\right\}$. 
For sufficiently small $N_n(1)$, $c_\alpha(N_n(1))$ equals the largest point in
the support of $B_\tau(N_n(1))$. A test that rejects only when the statistic
strictly exceeds this critical value can then never reject. For
$\tau=0.5$ and $\alpha=0.05$, this occurs when $N_n(1)\leq 5$. For such
extremely small groups, the statistic cannot exceed the rejection
threshold, and the lower bound in Theorem~\ref{thm:power} is
correspondingly uninformative. 

This limitation is not particular to the proposed procedure. Information cannot be created when the
treated-profile moment has too few possible realizations. In
applications, obtaining power for such groups generally requires
additional restrictions.

Overall, the simulations provide three conclusions. First, holding
the expected treatment-group size fixed, the feasible test approaches
its oracle behavior as the reference group grows. Second, its null
rejection probability is largely insensitive to the shape and tails
of the disturbance distribution. Third, the procedure can have
substantial power with a small treatment group, but its power is
ultimately constrained by the information and discreteness inherent
in extremely sparse groups.

\section{Empirical application: The Effect of Weather on Growth}
\label{sec:application}

In this section we apply our method to gauge the effect of large temperature changes on the cross-sectional distribution of GDP growth. We use the data from \cite{dell2012temperature} and complement their analysis in two directions. First, we study the effect of temperature changes on the 0.25, 0.50 and 0.75  quantiles of GDP growth. Second, we construct treatment profiles that allow us to estimate a saturated model as opposed to the linear regression model used by \cite{dell2012temperature} and that was presented in Section \ref{exa:Canon}. The purpose of this exercise is to compare the conclusions we obtain from doing inference based on standard asymptotics to what we obtain when using the method proposed in our paper.

Let \(Y_{i,t}\) denote GDP growth in country \(i\) over 1990--1994,
where \(t=1994\), and define the five-period treatment profile $D_i^t=(D_{i,1990},\ldots,D_{i,1994})$.\footnote{We restrict the sample to countries that are below the median of the GDP distribution in 1950, since those are the countries for which \cite{dell2012temperature} find a significant effect of temperature changes on GDP.} 
For each \(s\in\{1990,\ldots,1994\}\), \(D_{i,s}=1\) if temperature in
country \(i\) increased by more than one degree Celsius between
\(s-1\) and \(s\). We take \(d^t[0]=0^t\) as the reference profile and
denote the observed nonreference profiles by
\(\mathcal D^t=\{d^t[1],\ldots,d^t[M]\}\).

For each \(\tau\in\{0.25,0.50,0.75\}\), we consider
\begin{align}
\mu_t(x_i^t,D^{t}_{i},\tau)
	&=
	\kappa(\tau)+\Lambda(\tau)'x_i^t
	+\sum_{m=1}^{M}\theta_m(\tau)
	1\{D_i^t=d^t[m]\},
	\label{eq:empirical-quantile}
\end{align}
where \(X_i^t\) contains region indicators and rainfall in 1994.
Thus, the profile-specific null hypothesis is $H_{0,m}(\tau):\theta_m(\tau)=0$, 
while the joint null of no effect across the observed nonreference
profiles is $H_0(\tau):\theta_m(\tau)=0$ for every $m\in\{1,\ldots,M\}$. 

For comparison, we also estimate by OLS the conditional-mean
specification
\begin{align}\label{eq:DJO} 
E[Y_{i,t}\mid X_i^t=x_i^t,D_i^t]
=
\kappa+\Lambda'x_i^t
+\sum_{m=1}^{M}\theta_m^{\mathrm{mean}}
1\{D_i^t=d^t[m]\}
\end{align}

We choose this application as suitable to illustrate our method since weather shocks are exogenous to country characteristics. This is important for plausibility of assumption \ref{ass:U-DGP}, that requires the $\tau$-th quantile of the CDF of $U(d^t)$ given $X^t$ to be the same for each treatment profile. In practice, this is akin to a selection on observables assumption, which means that temperature shocks should be uncorrelated with unobserved factors conditional on region and baseline rainfall.

\begin{table}[h!]
	\centering
	\caption{Number of observations for each treatment profile}
	\begin{tabular}{lcc} \hline
		profile&No. \\
\hline
00000&49 \\
00001&6 \\
00011&1 \\
00100&1 \\
10000&4 \\
Total&61 \\ \hline
 
	\end{tabular}
	\footnotetext{Notes: We use the data from \cite{dell2012temperature} and restrict our attention to countries that were below the median of GDP in 1950. We then look at temperature shocks from 1990-1994 under the definition that a temperature shock is a one-year increase in temperature of one degree Celsius. We put these shocks together into treatment profiles.}
	\label{tab:DJO.Groups}
\end{table}

In Table \ref{tab:DJO.Groups} we show the size of each treatment group defined by $(D_{1990},...,D_{1994})$. We can see that the control group, $00000$, has 49 units whereas the other treatment groups have fewer observations, oscillating between 1 to 6 observations. Hence, standard inferential theory based on t-statistics or F-statistics can be unreliable and untrustworthy.  For profiles \(00011\) and \(00100\), the sample contains only one
country. We do not apply our procedure to these profiles because the
corresponding empirical quantile moments have too few attainable
values to yield informative inference. For the same reason, the OLS
and conventional quantile-regression results for these profiles should
not be viewed as reliable: a profile-specific coefficient based on a
single observation contains essentially no information about its
sampling variability.

We report the estimated coefficients in Tables \ref{tab:DJO_result} and \ref{tab:DJO_result_test}. In column 1 of Table \ref{tab:DJO_result} we show that, when estimating the coefficients in equation \ref{eq:DJO} by OLS and using standard errors robust to heteroskedasticity, only contemporaneous temperature seems to matter for growth, i.e., the profile $00001$. The same result holds for quantiles 0.25, 0.5, and 0.75, which we show in columns 2-4. However, as shown in Table \ref{tab:DJO_result_test}, when using our method to test the null hypothesis that $\theta_{00001} = 0$, we cannot reject it at a $95\%$ confidence level. Moreover, when testing the joint null hypothesis of no effect from  any temperature shock during the 1990-1994 period, i.e., $\theta_{00001} = \theta_{00011} = ... = \theta_{10000} = 0$, we also cannot reject this null.\footnote{The test statistic we use to test $\theta_{00001} = 0$ is analogous to the one used in the Monte Carlo study. The test statistic we use to test $\theta_{00001} = \theta_{00011} = ... = \theta_{10000} = 0$ is also analogous, but we constructed the treated moment by pooling the observations for all the profiles (other than $00000$).} 

The distinction is economically consequential. The point estimates for
profile $00001$ are large relative to mean growth in the control group,
so conventional inference suggests that the corresponding temperature
shock substantially reduces growth. However, this conclusion rests on
a treatment-profile group containing only six countries. Our procedure
does not reject the absence of an effect, indicating that the statistical
evidence for these estimated growth losses is sensitive to how inference
accounts for the small treatment group. This does not establish that
temperature shocks have no effect on growth. Rather, the comparison
illustrates how conventional significance assessments can overstate the
strength of the evidence for economically important effects when the
number of observations informative about those effects is small.

{\small{
\begin{table}[h!]
	\centering
	{
\def\sym#1{\ifmmode^{#1}\else\(^{#1}\)\fi}
\begin{tabular}{l*{4}{c}}
\hline\hline
                    &\multicolumn{1}{c}{(1)}&\multicolumn{1}{c}{(2)}&\multicolumn{1}{c}{(3)}&\multicolumn{1}{c}{(4)}\\
                    &\multicolumn{1}{c}{OLS}&\multicolumn{1}{c}{$\tau=0.25$}&\multicolumn{1}{c}{$\tau=0.5$}&\multicolumn{1}{c}{$\tau=0.75$}\\
\multicolumn{4}{l}{Dependent variable: GDP growth} \\ \hline
profile=00001      &        -7.4\sym{***}&        -7.7\sym{***}&         -11\sym{***}&        -6.4\sym{***}\\
                    &       (2.4)         &       (1.5)         &       (2.9)         &       (1.6)         \\
[1em]
profile=10000      &         5.2         &         3.6         &         4.8\sym{**} &         1.7         \\
                   &       (3.5)         &       (3.2)         &       (1.9)         &       (2.6)         \\
\hline
Moment of dependent variable in the 00000 group&        -.82         &       -2.82         &         2.1         &        4.99         \\
Observations        &          60         &          60         &          60         &          60         \\
\hline\hline
\end{tabular}
}

	\caption{Estimates of the effect of different profiles of weather shocks on output growth. }
	\footnotetext{Notes: These estimates use the same data from \cite{dell2012temperature} as in Table \ref{tab:DJO.Groups}. We restrict our attention to countries that were below the median of GDP in 1950. We then look at temperature shocks from 1990-1994 under the definition that a temperature shock is a one-year increase in temperature of one degree Celsius, putting these shocks together into treatment profiles. See Table \ref{tab:DJO.Groups} for the number of countries in each treatment profile.}
		\label{tab:DJO_result} 
\end{table}
}}

\begin{table}[h!]
	\centering
\begin{tabular}{lccccc} \hline\hline
    $H_{0} \setminus \text{Inference Method}$ &  (1) & (2) & (3) & (4) & \text{Our Method}  \\ \hline \hline 
    $\theta_{00001} = 0 $              & Reject & Reject & Reject & Reject & Do~not~reject \\
    $\theta_{00001} = ... = \theta_{10000} = 0$ & Reject & Reject & Reject & Reject &  Do~not~reject \\ \hline\hline
\end{tabular}
	\caption{Hypothesis test at 95\% confidence level.}
 \footnotetext{Notes: These estimates use the same data from \cite{dell2012temperature} as in Table \ref{tab:DJO.Groups}. We restrict our attention to countries that were below the median of GDP in 1950. We then look at temperature shocks from 1990-1994 under the definition that a temperature shock is a one-year increase in temperature of one degree Celsius, putting these shocks together into treatment profiles. See Table \ref{tab:DJO.Groups} for reference of how many countries have each treatment profile.}
		\label{tab:DJO_result_test} 
\end{table}

\pagebreak

\bibliography{ReferencesMDDP}
\pagebreak

\appendix 

\setcounter{page}{1}

\begin{center}
{\Huge{Online Appendix}}
\end{center}

\section{Proof of Proposition \ref{prop:structural-quantile}}
\label{app:structural.QTE}

\begin{proof}
	Fix $(t,d^t,y^{t-1},x^t)$. By expression 
	\eqref{eqn:PO-process} and Assumption
	\ref{ass:monotonicity}, $\left\{
		Y_{i,t}(d^t)
		\leq
		\mu_t(y^{t-1},x^t,d^t,\tau)
		\right\}  =
		\left\{
		U_{i,t}(d^t)\leq\tau
		\right\}$ 
	on the event $\left\{
	Y_i^{t-1}(d^{t-1})=y^{t-1},
	X_i^t=x^t
	\right\}$.
	Assumption \ref{ass:U-DGP} and the law of iterated expectations
	therefore imply
	\begin{align*}
		 P\left(
		Y_{i,t}(d^t)
		\leq
		\mu_t(Y_i^{t-1}(d^{t-1}),X^t_{i},d^t,\tau) \mid
		Y_i^{t-1}(d^{t-1}),
		X_i^t
		\right)  =
		P\left(
		U_{i,t}(d^t)\leq\tau
		\,\middle|\,
		Y_i^{t-1}(d^{t-1}),
		X_i^t
		\right)  =
		\tau.
	\end{align*}
	Continuity and strict monotonicity at $\tau$ imply that this
	conditional quantile is unique, establishing
	\eqref{eqn:structural-quantile}. Equation \eqref{eqn:observed-quantile-moment} then follows directly
	from Assumptions \ref{ass:monotonicity} and \ref{ass:U-DGP}.
\end{proof}

\section{Appendix for Section \ref{sec:main}}

\subsection{Proof of Lemma \ref{lem:conditional-bernoulli}}
\label{app:conditional-bernoulli}

\begin{proof}
	For \(m\in\{0,\ldots,M\}\), write $\mathcal G_{n,m}
	:=
	\mathcal G_n\bigl(x^t,d^t[m]\bigr)$ and $N_{n,m}:=
	N_n\bigl(x^t,d^t[m]\bigr)$. 	Fix \(i\in\mathcal G_{n,m}\). 
	
Since
	\[
	Y_{i,t}
	=
	Y_{i,t}\bigl(d^t[m]\bigr)
	=
	\mu\!\left(
	Y_i^{t-1},x^t,d^t[m],
	U_{i,t}\bigl(d^t[m]\bigr)
	\right),
	\]
	and under \(H_0\), $\mu_0(\cdot)
	=
	\mu\bigl(\cdot,x^t,d^t[m],\tau\bigr)$, 
	for every \(m\in\{0,\ldots,M\}\), Assumption \ref{ass:monotonicity} implies
	\begin{align}
		\rho\bigl(Y_{i,t},\mu_0(Y_i^{t-1})\bigr)
		&=
		1\left\{
		\mu\!\left(
		Y_i^{t-1},x^t,d^t[m],
		U_{i,t}\bigl(d^t[m]\bigr)
		\right)
		\leq
		\mu\!\left(
		Y_i^{t-1},x^t,d^t[m],\tau
		\right)
		\right\} =
		1\left\{
		U_{i,t}\bigl(d^t[m]\bigr)\leq\tau
		\right\}.
		\label{eq:proof-rank-indicator}
	\end{align}
	Define $Z_{i,m}
	:=
	1\left\{
	U_{i,t}\bigl(d^t[m]\bigr)\leq\tau  \right\}$, for all $i\in\mathcal G_{n,m}$. 
	On the event \(i\in\mathcal G_{n,m}\),
	Assumption \ref{ass:U-DGP} gives
	\begin{align*}
		\Pr\left(
		Z_{i,m}=1
		\,\middle|\,
		O_i^t
		\right)
		=
		\Pr\left(
		U_{i,t}\bigl(d^t[m]\bigr)\leq\tau
		\,\middle|\,
		Y_i^{t-1},X_i^t=x^t,D_i^t=d^t[m]
		\right) =\tau
		\qquad\text{a.s.}
	\end{align*}
	Thus, conditionally on \(O_i^t\), \(Z_{i,m}\) is Bernoulli with success
	probability \(\tau\).
	
	Assumption \ref{ass:DGP} implies that the complete unit-level random vectors are
	independent across \(i\). Consequently, conditionally on
	\((O_i^t)_{i=1}^n\), the variables $\left\{
	Z_{i,m}:
	m\in\{0,\ldots,M\},\ i\in\mathcal G_{n,m}
	\right\}$ 
	are mutually independent. Notice that there is at most one such variable
	for each \(i\), because the profile groups \(\mathcal G_{n,m}\) are
	pairwise disjoint. Their joint conditional distribution is therefore the
	same as that of the corresponding subcollection of
	\(B_1,\ldots,B_n\), where the \(B_i\)'s are IID
	\(\operatorname{Bernoulli}(\tau)\) variables independent of
	\((O_i^t)_{i=1}^n\).
	
	Using \eqref{eq:proof-rank-indicator}, for every \(m\),
	$\widehat m_{n,K_m}(x^t,d^t[m],\mu_0)
		=
		\frac{1}{N_{n,m}}
		\sum_{i\in\mathcal G_{n,m}}
		\left(Z_{i,m}-\tau\right)
		\Phi_{K_m}(Y_i^{t-1})$. 
	Conditionally on \((O_i^t)_{i=1}^n\), the group memberships, group sizes,
	and vectors \(\Phi_{K_m}(Y_i^{t-1})\) are fixed. It follows from the
	joint conditional Bernoulli law established above that
	\[
	\left(
	\widehat m_{n,K_m}(x^t,d^t[m],\mu_0)
	\right)_{m=0}^M
	\overset{d}{=}
	\left(
	\frac{1}{N_{n,m}}
	\sum_{i\in\mathcal G_{n,m}}
	(B_i-\tau)\Phi_{K_m}(Y_i^{t-1})
	\right)_{m=0}^M
	=
	\left(
	\ell_{n,K_m}(d^t[m])
	\right)_{m=0}^M
	\]
	conditionally on \((O_i^t)_{i=1}^n\). This proves
	\eqref{eq:joint-bernoulli-representation}.
	
	Finally, \(T_{n,\mathbf K}^0\) and \(T_{n,\mathbf K}^B\) are obtained by
	applying the same measurable aggregation rule to the two respective
	vectors of moments. The preceding joint conditional distributional
	equality therefore implies $T_{n,\mathbf K}^0 \mid 
	(O_i^t)_{i=1}^n
	\overset{d}{=}
	T_{n,\mathbf K}^B \mid 
	(O_i^t)_{i=1}^n$, 
	which proves \eqref{eq:oracle-statistic-bernoulli-law}.
\end{proof}

\subsection{Proof of Theorem \ref{thm:feasible-validity}} 
\label{app:proof.size}

We establish the result of Theorem \ref{thm:feasible-validity} under the more general specification of 
\[
T_{n,\mathbf K}(h)
=
H\!\left(
\widehat m_{n,K_0}(x^t,d^t[0],h),\ldots,
\widehat m_{n,K_M}(x^t,d^t[M],h)
\right),
\]
where \(H(0,\ldots,0)=0\) and, for some finite \(C\), $\left|
H(a_0,\ldots,a_M)-H(b_0,\ldots,b_M)
\right|
\leq
C\max_{0\leq m\leq M}
\left|
\|a_m\|_2^2-\|b_m\|_2^2
\right|$. 
This class includes the sum and maximum statistics, as well as
weighted versions with uniformly bounded weights measurable with
respect to \((O_i^t)_{i=1}^n\).

\begin{proof}[Proof of Theorem~\ref{thm:feasible-validity}]
	Let $\mathcal O_n:=\sigma\!\left((O_i^t)_{i=1}^n\right)$ be the $\sigma$-algebra generated by $(O_i^t)_{i=1}^n$. Also, let $T_{n,\mathbf K(n)}^0
	:=
	T_{n,\mathbf K(n)}(\mu_0)$. 
	Under \(H_0\), Lemma~\ref{lem:conditional-bernoulli} gives
	\[
	\left(
	\widehat m_{n,K_m(n)}
	(x^t,d^t[m],\mu_0)
	\right)_{m=0}^M
	\ \bigg|\ \mathcal O_n
	\overset{d}{=}
	\left(
	\ell_{n,K_m(n)}(d^t[m])
	\right)_{m=0}^M
	\ \bigg|\ \mathcal O_n .
	\]
	Since \(H\) is measurable and any data-dependent components of \(H\),
	such as weights, are \(\mathcal O_n\)-measurable, the conditional
	mapping theorem implies $T_{n,\mathbf K(n)}^0\mid\mathcal O_n
	\overset{d}{=}
	T_{n,\mathbf K(n)}^B\mid\mathcal O_n$, 
	where
	\[
	T_{n,\mathbf K(n)}^B
	=
	H\!\left(
	\ell_{n,K_0(n)}(d^t[0]),\ldots,
	\ell_{n,K_M(n)}(d^t[M])
	\right).
	\]
	Consequently, by the definition of the conditional critical value, $P\!\left(
	T_{n,\mathbf K(n)}^0
	>
	c_{n,\mathbf K(n)}(\alpha)
	\,\middle|\,
	\mathcal O_n
	\right)
	\leq \alpha$ a.s. Taking expectations yields $P\!\left(
		T_{n,\mathbf K(n)}^0
		>
		c_{n,\mathbf K(n)}(\alpha)
		\right)
		\leq \alpha$. 
	
	Now define $\mathcal A_n
	:=
	\left\{
	\widehat T_{n,\mathbf K(n)}
	-
	T_{n,\mathbf K(n)}^0
	\leq \epsilon_n
	\right\}$. On \(\mathcal A_n\), $\widehat T_{n,\mathbf K(n)}
	\leq
	T_{n,\mathbf K(n)}^0+\epsilon_n$. 
	Therefore, $\left\{
	\widehat T_{n,\mathbf K(n)}
	>
	c_{n,\mathbf K(n)}(\alpha)+\epsilon_n
	\right\}
	\cap\mathcal A_n
	\subseteq
	\left\{
	T_{n,\mathbf K(n)}^0
	>
	c_{n,\mathbf K(n)}(\alpha)
	\right\}$. It follows that
	\begin{align*}
		E_P\!\left[\varphi_{n,\mathbf K(n)}(\alpha)\right]
		&=
		P\!\left(
		\widehat T_{n,\mathbf K(n)}
		>
		c_{n,\mathbf K(n)}(\alpha)+\epsilon_n
		\right)\\
		&\leq
		P\!\left(
		T_{n,\mathbf K(n)}^0
		>
		c_{n,\mathbf K(n)}(\alpha)
		\right)
		+
		P(\mathcal A_n^c)\\
		&\leq
		\alpha+
		P\!\left(
		\widehat T_{n,\mathbf K(n)}
		-
		T_{n,\mathbf K(n)}^0
		>
		\epsilon_n
		\right).
	\end{align*}
	The final probability converges to zero by
	Assumption~\ref{ass:oracle-approximation}. Hence, $\limsup_{n\to\infty}
	E_P\!\left[\varphi_{n,\mathbf K(n)}(\alpha)\right]
	\leq\alpha$, 
	as claimed.
\end{proof}

\subsection{Proof of Theorem \ref{thm:power}} 
\label{app:power}

\begin{proof}[Proof of Theorem \ref{thm:power}]
	For notational simplicity, let
	$
	\mathcal G_{n,l}:= \mathcal G_n(x^t,d^t[l])$, $N_{n,l}:=N_n(x^t,d^t[l])$, and $\mu_{i,l} : = \mu(Y^{t-1}_{i},x^{t},d^{t}[l],\tau)$.
	
	Let $P_{t}$ denote the probability measure conditional on $(O^{t}_{i})_{i=1}^{n}$, and let $E_{t}$, $\operatorname{Var}_{t}$ denote its associated moments.

	Let $R_{n} : = P_{t}( |\widehat{T}_{n,\mathbf{K}(n)} - T^{0}_{n,\mathbf{K}(n)}| > \epsilon_{n} )$. Under Assumption \ref{ass:oracle-approximation}, Law of Iterated Expectations, and the Markov inequality. $R_{n} = o_{P}(1)$. Hence,
	\[
	\begin{split}
		P_{t}\left(
		\widehat{T}_{n,\mathbf{K}(n)} \geq c_{n,K(n)}(\alpha)+\epsilon_{n}
		\right)
		\geq &
		P_{t}\left(
		T^{0}_{n,\mathbf{K}(n)} \geq c_{n,K(n)}(\alpha)+3\epsilon_{n}
		\right) -
		P_{t}\left(
		\left|\widehat{T}_{n,\mathbf{K}(n)} - T^{0}_{n,\mathbf{K}(n)} \right|\geq 2\epsilon_{n}
		\right)\\
		\geq & 
		P_{t}\left(
		T^{0}_{n,\mathbf{K}(n)} \geq c_{n,K(n)}(\alpha)+3\epsilon_{n}
		\right) - o_{P}(1).
	\end{split}
	\]
	It therefore suffices to obtain a lower bound for the first
	probability on the right-hand side.

	Observe that for any specification of $T^{0}$, 
	\begin{align*}
		T^{0}_{n,\mathbf{K}(n)} \geq ||	\widehat{m}_{n,K_{n}}(x^{t},d^{t}[l],\mu_{0})||^{2},~\forall l \in \{0,...,M\}.
	\end{align*}
	Thus, 	for all $l \in \{0,...,M\}$, 
	\begin{align*}
		P_{t} \left( 	T^{0}_{n,\mathbf{K}(n)} \geq c_{n,K(n)}(\alpha)+3\epsilon_{n}  \right)  \ge 	P_{t} \left( ||	\widehat{m}_{n,K_{n}}(x^{t},d^{t}[l],\mu_{0})||^{2} \geq c_{n,K(n)}(\alpha)+3\epsilon_{n}  \right) .
	\end{align*}

	Let $\Phi_i:=\Phi_{K_l}(Y_i^{t-1})$, and  $p_{i,n}
	:=
	F\left(
	\mu(0)
	\mid
	Y_i^{t-1},X_i^t,D_i^t=d^t[l]
	\right)$. 	Conditionally on \(O^{t}_{i}\), $B_{i,n}:=1\{Y_{i,t}\leq\mu_{i,0} \}$, for all	$i \in\mathcal G_{n,l}$, 
	are independent Bernoulli random variables with success probabilities
	\(p_{i,n}\).
	
	Define
	\begin{align*}
		a_{n,l}
		:=
		\frac{1}{N_{n,l}}
		\sum_{i\in\mathcal G_{n,l}}
		(p_{i,n}-\tau)\Phi_i, \quad and \quad
		\zeta_{n,l}
		:=
		\frac{1}{N_{n,l}}
		\sum_{i\in\mathcal G_{n,l}}
		(B_{i,n}-p_{i,n})\Phi_i.
	\end{align*}
	Thus, the following decomposition follows $\widehat{m}_{n,K_{n}}(x^{t},d^{t}[l],\mu_{0})
	=
	a_{n,l}+\zeta_{n,l}$. 
	Let	$h_{n,l}:=\sqrt{N_{n,l}}\,a_{n,l}$ and $Z_{n,l}:=\sqrt{N_{n,l}}\,\zeta_{n,l}$,  so that
	\begin{equation}
		N_{n,l} \left\|  \widehat{m}_{n,K_{n}}(x^{t},d^{t}[l],\mu_{0}) \right\|^2
		=
		\|h_{n,l}+Z_{n,l}\|^2.
		\label{eq:normalized-oracle-statistic}
	\end{equation}
	
	Suppose that \(h_{n,l}\neq0\), and define $u_{n,l}
	:=
	\frac{h_{n,l}}{\|h_{n,l}\|}$. 
	Let
	\begin{align*}
		\sigma_{n,l}^2
		&:=
		\frac{1}{N_{n,l}}
		\sum_{i\in\mathcal G_{n,l}}
		p_{i,n}(1-p_{i,n})
		\bigl(u_{n,l}'\Phi_i\bigr)^2,~and~
		\rho_{n,l}
		&:=
		\frac{
			N_{n,l}^{-3/2}
			\displaystyle\sum_{i\in\mathcal G_{n,l}}
			E_t\left[
			|B_{i,n}-p_{i,n}|^3
			\right]
			|u_{n,l}'\Phi_i|^3
		}{
			\sigma_{n,l}^3
		}.
	\end{align*}

	We now show that, for every \(c_{n,l}\geq0\),
	\begin{align} \notag
		&Pr_t\left(
		\left\|   \widehat{m}_{n,K_{n}}(x^{t},d^{t}[l],\mu_{0})   \right\|^2
		>
		c_{n,l}
		\right) \\
		&\quad\geq
		1-
			\operatorname{F}_{\mathcal{N}(0,1)}\left(
		\frac{
			\sqrt{N_{n,l}c_{n,l}}-\|h_{n,l}\|
		}{
			\sigma_{n,l}
		}
		\right)
		+
			\operatorname{F}_{\mathcal{N}(0,1)} \left(
		\frac{
			-\sqrt{N_{n,l}c_{n,l}}-\|h_{n,l}\|
		}{
			\sigma_{n,l}
		}
		\right)
		-
		2C_{\mathrm{BE}}\rho_{n,l},
		\label{eq:berry-esseen-power-bound}
	\end{align}
	where $	\operatorname{F}_{\mathcal{N}(0,1)}$ denotes the standard normal distribution function and
	\(C_{\mathrm{BE}}\) is a universal Berry--Esseen constant.
	
	By definition	$
	Z_{n,l}
	=
	\frac{1}{\sqrt{N_{n,l}}}
	\sum_{i\in\mathcal G_{n,l}}
	(B_{i,n}-p_{i,n})\Phi_i$. 	Consider its projection in the direction of the conditional mean: $u_{n,l}'Z_{n,l}
	=
	\frac{1}{\sqrt{N_{n,l}}}
	\sum_{i\in\mathcal G_{n,l}}
	(B_{i,n}-p_{i,n})u_{n,l}'\Phi_i$. 
	Since \(\|u_{n,l}\|=1\), the Cauchy--Schwarz inequality gives $\|h_{n,l}+Z_{n,l}\|
	\geq
	\left|
	u_{n,l}'(h_{n,l}+Z_{n,l})
	\right|$. 
	Moreover, $u_{n,l}'h_{n,l}
	=
	\frac{h_{n,l}'h_{n,l}}{\|h_{n,l}\|}
	=
	\|h_{n,l}\|$. 
	It follows that $\|h_{n,l}+Z_{n,l}\|
		\geq
		\left|
		\|h_{n,l}\|+u_{n,l}'Z_{n,l}
		\right|$. Consequently,
	\begin{align*}
		&
		\left\{
		\left|
		\|h_{n,l}\|+u_{n,l}'Z_{n,l}
		\right|^2
		>
		N_{n,l}c_{n,l}
		\right\} \subseteq
		\left\{
		\|h_{n,l}+Z_{n,l}\|^2
		>
		N_{n,l} c_{n,l}
		\right\}.
	\end{align*}

	Combining this inclusion with
	\eqref{eq:normalized-oracle-statistic} yields
	\begin{align}
		&Pr_t\left(
		\left\|     \widehat{m}_{n,K_{n}}(x^{t},d^{t}[l],\mu_{0})     \right\|^2
		>
		c_{n,l}
		\right) \geq
		Pr_t\left(
		\left|
		\|h_{n,l}\|+u_{n,l}'Z_{n,l}
		\right|
		>
		\sqrt{N_{n,l} c_{n,l}}
		\right).
		\label{eq:projection-power-bound}
	\end{align}
	
	We next approximate the conditional distribution of
	\(u_{n,l}'Z_{n,l}\). Define $W_{n,l}
	:=
	\frac{u_{n,l}'Z_{n,l}}{\sigma_{n,l}}
	=
	\sum_{i\in\mathcal G_{n,l}}X_{i,n}$, 
	where $X_{i,n}
	:=
	\frac{
		(B_{i,n}-p_{i,n})u_{n,l}'\Phi_i
	}{
		\sqrt{N}\,\sigma_{n,l}
	}$. Let	$r_{n,l}:=\sqrt{N_{n,l} c_{n,l}}$ and $
	d_{n,l}:=\|h_{n,l}\|$.  The two events $\left\{
	u_{n,l}'Z_{n,l}>r_{n,l}-d_{n,l}
	\right\}$ and $\left\{
	u_{n,l}'Z_{n,l}<-r_{n,l}-d_{n,l}
	\right\}$ 
	are disjoint. Hence,
	\begin{align}
		Pr_t\left(
		|d_{n,l}+u_{n,l}'Z_{n,l}|>r_{n,l}
		\right) =
		Pr_t\left(
		W_{n,l}>
		\frac{r_{n,l}-d_{n,l}}{\sigma_{n,l}}
		\right)
		+
		Pr_t\left(
		W_{n,l}<
		\frac{-r_{n,l}-d_{n,l}}{\sigma_{n,l}}
		\right).
		\label{eq:two-tail-decomposition}
	\end{align}

	Conditionally on \( ( O^{t}_{i} )_{i}  \), the variables
	\(\{X_{i,n}:i\in\mathcal G_{n,l}\}\) are independent and satisfy $E_t[X_{i,n}]=0$. Furthermore,
	\begin{align*}
		\sum_{i\in\mathcal G_{n,l}}E_t[X_{i,n}^2]
		=
		\frac{
			N_{n,l}^{-1}
			\displaystyle\sum_{i\in\mathcal G_{n,l}}
			p_{i,n}(1-p_{i,n})
			(u_{n,l}'\Phi_i)^2
		}{
			\sigma_{n,l}^2
		} = 1.
	\end{align*}
	Their aggregate third absolute moment is
	\begin{align*}
		\sum_{i\in\mathcal G_{n,l}}E_t[|X_{i,n}|^3]
		=
		\frac{
			N_{n,l}^{-3/2}
			\displaystyle\sum_{i\in\mathcal G_{n,l}}
			E_t\left[
			|B_{i,n}-p_{i,n}|^3
			\right]
			|u_{n,l}'\Phi_i|^3
		}{
			\sigma_{n,l}^3
		} =\rho_{n,l}.
	\end{align*}
	The Berry--Esseen inequality therefore implies
	\begin{equation}
		\sup_{z\in\mathbb R}
		\left|
		\Pr_t(W_{n,l}\leq z)-\Phi(z)
		\right|
		\leq
		C_{\mathrm{BE}}\rho_{n,l}.
		\label{eq:conditional-berry-esseen}
	\end{equation}

	By \eqref{eq:conditional-berry-esseen},
	\begin{align*}
		Pr_t\left(
		W_{n,l}>
		\frac{r_{n,l}-d_{n,l}}{\sigma_{n,l}}
		\right)
		&\geq
		1-
		\operatorname{F}_{\mathcal N(0,1)}\left(
		\frac{r_{n,l}-d_{n,l}}{\sigma_{n,l}}
		\right)
		-
		C_{\mathrm{BE}}\rho_{n,l},\\
		Pr_t\left(
		W_{n,l}<
		\frac{-r_{n,l}-d_{n,l}}{\sigma_{n,l}}
		\right)
		&\geq
		\operatorname{F}_{\mathcal N(0,1)}\left(
		\frac{-r_{n,l}-d_{n,l}}{\sigma_{n,l}}
		\right)
		-
		C_{\mathrm{BE}}\rho_{n,l}.
	\end{align*}
	Combining these two inequalities with
	\eqref{eq:projection-power-bound} and
	\eqref{eq:two-tail-decomposition} gives
	\eqref{eq:berry-esseen-power-bound}.

	We now refine the Berry-Esseen rate $\rho_{n,l}$. Observe that 
	\begin{align*}
		E_t\left[
		|B_{i,n}-p_{i,n}|^3
		\right] =
		p_{i,n}(1-p_{i,n})^3
		+
		(1-p_{i,n})p_{i,n}^3 =
		p_{i,n}(1-p_{i,n})
		\left\{
		p_{i,n}^2+(1-p_{i,n})^2
		\right\} \leq \frac14.
	\end{align*}
	
	Therefore, 
	\begin{align*}
		\rho_{n,l} 	\leq \frac14
		\frac{
			N_{n,l}^{-3/2}
			\displaystyle\sum_{i\in\mathcal G_{n,l}} 
			|u_{n,l}'\Phi_i|^3
		}{
			\sigma_{n,l}^3
		} \leq  \frac14
		\frac{
			N_{n,l}^{-3/2}
			\displaystyle\sum_{i\in\mathcal G_{n,l}} 
			||\Phi_i||^3
		}{
			\sigma_{n,l}^3
		}  \leq  \frac14
		\frac{
			N_{n,l}^{-1/2} B_{K_l(n)}^{3}
		}{
			\sigma_{n,l}^3
		}
	\end{align*}
	where $\sup_y\left\|\Phi_{K_l(n)}(y)\right\|
	\leq B_{K_l(n)}$.

	We now bound $||h_{n,l}||$. By Proposition \ref{prop:structural-quantile}, $\tau = F(\mu_{i,l} \mid Y^{t-1}_{i},X^{t}_{i},D^{t}_{i} = d^{t}[l] )$. Hence,
	\begin{align*}
		p_{i,n} - \tau = F(\mu_{i,0} \mid Y^{t-1}_{i},X^{t}_{i},D^{t}_{i} = d^{t}[l] ) - F(\mu_{i,l} \mid Y^{t-1}_{i},X^{t}_{i},D^{t}_{i} = d^{t}[l] ))
	\end{align*}
	and by absolute continuity of $F(\cdot \mid Y^{t-1}_{i},X^{t}_{i},D^{t}_{i} = d^{t}[l] )$,
	\begin{align*}
		\left|  F(\mu_{i,0} \mid Y^{t-1}_{i},X^{t}_{i},D^{t}_{i} = d^{t}[l] ) - F(\mu_{i,l} \mid Y^{t-1}_{i},X^{t}_{i},D^{t}_{i} = d^{t}[l] ) \right|  \geq \underbar{f}_{l} \times | \Delta_{l}(Y^{t-1}_{i},X^{t}_{i})|.
	\end{align*}
	
	Since the first component of $\Phi_{K}(y)$ is assumed to be one, $e'_{1} \Phi_{K}(y) = 1$ where $e'_{1} : = (1,0,...,0)$. Moreover, $||	a_{n,l} || \geq |e_{1}' a_{n,l} |$. 
	
	This observation and the last two displays imply
	\begin{align*}
		||	a_{n,l} || \geq & \left|  \frac{1}{N_{n,l}} \sum_{i \in \mathcal{G}_{n,l}  } (F(\mu_{i,0} \mid Y^{t-1}_{i},X^{t}_{i},D^{t}_{i} = d^{t}[l] ) - F(\mu_{i,l} \mid Y^{t-1}_{i},X^{t}_{i},D^{t}_{i} = d^{t}[l] ))   \right|  \\
		\geq & \underbar{f}_{l}   \frac{1}{N_{n,l}} \sum_{i \in \mathcal{G}_{n,l}  } | \Delta_{l}(Y^{t-1}_{i},X^{t}_{i})|.
	\end{align*}	
	Hence, $||h_{n,l}|| \geq \underbar{f}_{l}   \frac{1}{\sqrt{ N_{n,l} } } \sum_{i \in \mathcal{G}_{n,l}  } | \Delta_{l}(Y^{t-1}_{i},X^{t}_{i})|$. 
	
	By inputting this lower bound and the upper bound on $\rho_{n,l}$ in expression \eqref{eq:berry-esseen-power-bound} it follows that 
	\begin{align*}
		Pr_t\left(
		\left\|   \widehat{m}_{n,K_{n}}(x^{t},d^{t}[l],\mu_{0})   \right\|^2
		>
		c_{n,l}
		\right)
		\geq &
		1-
			\operatorname{F}_{\mathcal{N}(0,1)}\left(
		\frac{
			\sqrt{N_{n,l}c_{n,l}}- \underbar{f}_{l}   \frac{1}{\sqrt{ N_{n,l} } } \sum_{i \in \mathcal{G}_{n,l}  } | \Delta_{l}(Y^{t-1}_{i},X^{t}_{i})| 
		}{
			\sigma_{n,l}
		}
		\right)\\
		& +
			\operatorname{F}_{\mathcal{N}(0,1)}\left(
		\frac{
			-\sqrt{N_{n,l}c_{n,l}}- \underbar{f}_{l}   \frac{1}{\sqrt{ N_{n,l} } } \sum_{i \in \mathcal{G}_{n,l}  } | \Delta_{l}(Y^{t-1}_{i},X^{t}_{i})| 
		}{
			\sigma_{n,l}
		}
		\right)\\
		& - 
		2C_{\mathrm{BE}}   \frac14
		\frac{
			N_{n,l}^{-1/2} B_{K_l(n)}^{3}
		}{
			\sigma_{n,l}^3
		}
	\end{align*}
	
	Since the second term in the RHS is positive, the lower bound in the statement of the theorem follows from this (stronger) inequality.

	%
	%

\end{proof}

\subsection{Proofs of Propositions  \ref{prop:oracle-location-scale} and \ref{prop:oracle-nonseparable}}
\label{app:Suff.Oracle.Examples}

\begin{proof}[Proof of Proposition \ref{prop:oracle-location-scale}]
	Because \(x^t\) is fixed, its contribution to the conditional location
	can be absorbed into the intercept. Define	$\alpha_m(x^{t}) :=\theta_{d^t[m]}+\lambda x^t$. 
	For profile \(d^t[m]\), the model can then be written as
	\[
	Y_t(d^t[m])
	=
	\alpha_m(x^{t})  +  \gamma Y_{t-1}(d^{t-1}[m])
	+
	\sigma_t(Y_{t-1},x^t,d^t[m])
	\epsilon_t(d^t[m]).
	\]
	The normalization $F_{d^t[m]}^{-1}(\tau\mid y_{t-1},x^t)=0$ implies that the reference-profile conditional \(\tau\)-quantile is $\mu_0(y^{t-1}): 
	=
	\alpha_0(x^{t}) +\gamma y_{t-1}$. 
	
	To obtain sufficient conditions for Assumption \ref{ass:oracle-approximation} to hold, we verify the sufficient conditions in
	Proposition \ref{prop:primitive-oracle-approximation} in Appendix \ref{app:oracle-sufficient-conditions}. For notational convenience,
	write $\sigma_m(y^{t-1})
	:=
	\sigma_t(y_{t-1},x^t,d^t[m])$. 
	Conditional on $Y^{t-1}=y^{t-1}$ and $(X^t,D^t)=(x^t,d^t[m])$, 
	the distribution function of \(Y_t\) is
	\[
	F_m(a\mid y^{t-1})
	=
	F_{\epsilon,d^t[m]}
	\left(
	\frac{a-\alpha_m-\gamma y_{t-1}}
	{\sigma_m(y^{t-1})}
	\,\middle|\,
	y^{t-1},x^t
	\right).
	\]
	Therefore, for any \(a,b\in\mathbb R\),
	\begin{align*}
		\left|
		F_m(a\mid y^{t-1})
		-
		F_m(b\mid y^{t-1})
		\right|
		\leq
		A
		\left|
		\frac{a-b}{\sigma_m(y^{t-1})}
		\right|   \leq
		\frac{A}{c}|a-b|.
	\end{align*}
	Thus, condition \eqref{eq:profile-cdf-holder} in Appendix \ref{app:oracle-sufficient-conditions} holds with $\delta=1$ and $A_0(y^{t-1})=\frac{A}{c}$. 
	Next, let $\beta_0:=(\alpha_0,\gamma)'$ and $\widehat\beta_n:=(\widehat\alpha_n,\widehat\gamma_n)'$. Since
	\[
	\widehat\mu_n(y^{t-1})-\mu_0(y^{t-1})
	= (1,y^{t-1})'	(\widehat\beta_n-\beta_0) = : 
	Z(y^{t-1})'
	(\widehat\beta_n-\beta_0),
	\]
	the Cauchy--Schwarz inequality gives, for every \(m\), $\left\|
	\widehat\mu_n-\mu_0
	\right\|_{L^2(Q(\cdot \mid x^{t},d^{t}[m]) )}
	\leq
	\|Z\|_{L^2(Q(\cdot \mid x^{t},d^{t}[m]) )}
	\|\widehat\beta_n-\beta_0\|_2$. 
	Root-\(n\) consistency of
	\(\widehat\beta_n\) therefore implies
	\[
	\max_{0\leq m\leq M}
	\left\|
	\widehat\mu_n-\mu_0
	\right\|_{L^2(Q(\cdot \mid x^{t},d^{t}[m]) )}
	=
	O_P(n^{-1/2}).
	\]
	Condition~\eqref{eq:reference-estimator-rate}  in Appendix \ref{app:oracle-sufficient-conditions} therefore holds with
	\(q=2\) and $r_n=n^{-1/2}$. 
	
	Since the moment dimensions are fixed and their \(L^2(Q(\cdot \mid x^{t},d^{t}[m]) )\) norms are
	uniformly bounded,
	\[
	b_n
	=
	\max_{0\leq m\leq M}
	\left\|
	\|\Phi_{K_m}\|_2
	\right\|_{L^2(Q(\cdot \mid x^{t},d^{t}[m]) )}
	=
	O(1).
	\]
	Moreover, because \(q'=2\) and the Hölder envelope is the constant
	\(A/c\),
	\begin{align*}
		a_n
		=
		r_n
		\max_{0\leq m\leq M}
		\left\|
		\frac{A}{c}\|\Phi_{K_m}\|_2
		\right\|_{L^2(Q(\cdot \mid x^{t},d^{t}[m]) )} =	O(n^{-1/2}).
	\end{align*}
	
	It remains to control the reference-group empirical-process term. In
	this example, the relevant class is
	\[
	\mathcal M_0
	=
	\left\{
	(y_t,y^{t-1})
	\mapsto
	\left[
	1\{y_t\leq\alpha+\gamma y_{t-1}\}
	-
	1\{y_t\leq\alpha_0+\gamma_0y_{t-1}\}
	\right]
	\Phi_{K_0}(y^{t-1})
	:
	(\alpha,\gamma)\in B
	\right\}.
	\]
	The indicator functions in this display form a fixed-dimensional
	VC-subgraph class. Because \(K_0\) is fixed and
	\(\|\Phi_{K_0}\|_2\) is a square-integrable envelope, the standard
	maximal inequality for VC-type classes gives
	\[
	\sup_{f\in\mathcal M_0}
	\left\|
	(P_{n,0}-P_0)f
	\right\|_2
	=
	O_P(N_{n,0}^{-1/2})
	=
	O_P(n^{-1/2}),
	\]
	where the last equality uses that the size of the reference group is
	proportional to \(n\). Hence, condition \ref{eq:reference-ulln}  in Appendix \ref{app:oracle-sufficient-conditions} holds with $s_n=n^{-1/2}$. 
	If the reference-profile moment is omitted from the statistic, this
	step is unnecessary and one may instead take \(s_n=0\).
	
	It follows that $\widetilde a_n:=a_n+s_n=O(n^{-1/2})$. 
	Combining this result with \(b_n=O(1)\) yields $b_n\widetilde a_n+\widetilde a_n^2
	=
	O(n^{-1/2})+O(n^{-1})
	=
	O(n^{-1/2})$.

	The rate condition in
	Proposition \ref{prop:primitive-oracle-approximation} is therefore satisfied.
	Consequently,
	\[
	P\left(
	\left|
	\widehat T_{n,\mathbf K}
	-
	T^0_{n,\mathbf K}
	\right|
	>
	\epsilon_n
	\right)
	\longrightarrow0,
	\]
	which establishes Assumption~\ref{ass:oracle-approximation} with $\epsilon_n=(\log\log n)n^{-1/2}$. 
\end{proof}

\begin{proof}[Proof of Proposition \ref{prop:oracle-nonseparable}]
	To establish the desired result we also use Proposition \ref{prop:primitive-oracle-approximation}  in Appendix \ref{app:oracle-sufficient-conditions}. Because \(U_t(d^t[m])\) is conditionally uniform and independent of
	the realized treatment profile conditional on the observed history,
	the conditional distribution function of \(Y_t\) in profile
	\(d^t[m]\) is $F_m(a\mid y^{t-1})
	=
	P\left(
	g_m(U,y^{t-1})\leq a
	\right)$,  $U\sim U(0,1)$. 
	Since \(u\mapsto g_m(u,y^{t-1})\) is strictly increasing, this can be
	written, over the range of \(g_m\), as $F_m(a\mid y^{t-1})
	=
	g_m^{-1}(a,y^{t-1})$, 
	where the inverse is taken with respect to the first argument.
	
	The lower-slope condition implies that the inverse is Lipschitz. In
	fact, if $u_a=g_m^{-1}(a,y^{t-1})$, and  $u_b=g_m^{-1}(b,y^{t-1})$, 
	then
	\[
	|a-b|
	=
	|g_m(u_a,y^{t-1})-g_m(u_b,y^{t-1})|
	\geq
	c_\mu|u_a-u_b|.
	\]
	Consequently,
	\[
	|F_m(a\mid y^{t-1})-F_m(b\mid y^{t-1})|
	\leq
	c_\mu^{-1}|a-b|.
	\]
	The same inequality continues to hold outside the range of \(g_m\)
	after truncating the inverse at zero and one. Thus,
	condition~\eqref{eq:profile-cdf-holder} holds with $\delta=1$ and $A(y^{t-1})=c_\mu^{-1}$. 
	
	By the domination condition,
	\begin{align*}
		\|\widehat\mu_{n,K_0(n)}-\mu_0\|_{L^2(Q(\cdot \mid x^{t},d^{t}[m]) )}^2
		& =
		\int
		|\widehat\mu_{n,K_0(n)}-\mu_0|^2
		\frac{dQ(\cdot \mid x^{t},d^{t}[m]) }{d P_{Y^{t-1}} }\,d P_{Y^{t-1}}                 \\
		&     \leq
		C
		\|\widehat\mu_{n,K_0(n)}-\mu_0\|_{L^2(P_{Y^{t-1}})}^2.
	\end{align*}
	Because the number of profiles is fixed,
	\[
	\max_{0\leq m\leq M}
	\|\widehat\mu_{n,K_0(n)}-\mu_0\|_{L^2(Q(\cdot \mid x^{t},d^{t}[m]) )}
	=
	O_P(r_n).
	\]
	Hence, condition~\eqref{eq:reference-estimator-rate} holds with
	\(q=2\) and $r_n=n^{-s/(2s+d)}$.

	Since the reference-profile moment is omitted, only the profiles
	\(m=1,\ldots,M\) enter the bounds below. By conditions 2 and 3, $b_n
	=
	\max_{1\leq m\leq M}
	\left\|\|\Phi_{K_m}\|_2\right\|_{L^2(Q(\cdot\mid x^t,d^t[m]))}
	\leq
	\sqrt{C}\,
	\max_{1\leq m\leq M}
	\left\|\|\Phi_{K_m}\|_2\right\|_{L^2(P_{Y^{t-1}})}
	=O(1)$. 
	Moreover, the H\"older envelope is the constant \(c_\mu^{-1}\),
	so, with \(\delta=1\) and \(q=q'=2\), $a_n=c_\mu^{-1}r_n b_n=O(r_n)$. 
	There is no reference-group empirical-process term, so \(s_n=0\)
	and \(\widetilde a_n=a_n\). Consequently, $b_n\widetilde a_n+\widetilde a_n^2
	=O(r_n+r_n^2)
	=O(r_n)
	=o(\epsilon_n)$, 
	where \(\epsilon_n=(\log\log n)r_n\).
	Proposition \ref{prop:primitive-oracle-approximation} therefore implies Assumption \ref{ass:oracle-approximation}.
	

\end{proof}

\section{Sufficient Conditions for Assumption \ref{ass:oracle-approximation}}
\label{app:oracle-sufficient-conditions}

We provide primitive conditions under which  Assumption \ref{ass:oracle-approximation} holds. The argument separates two effects of replacing the
reference-profile quantile function \(\mu_0\) with its estimator
\(\widehat\mu_{n,K_0(n)}\). The first is the population change in the
profile-specific moments. The second arises only for the reference-profile
moment: because the same observations are used to estimate \(\mu_0\) and to
evaluate that moment, a uniform law of large numbers over the reference-profile
sieve is also required.

Here and throughout, define the
empirical and population measures for the reference group by
\begin{align}
	P_{n,0}f
	&:=
	\frac{1}{N_n(x^t,d^t[0])}
	\sum_{i\in\mathcal G_n(x^t,d^t[0])}
	f(Y_{i,t},Y_i^{t-1}),
	\label{eq:reference-empirical-measure}\\
	P_0f
	&:=
	E\left[
	f(Y_t,Y^{t-1})
	\,\middle|\,
	(X^t,D^t)=(x^t,d^t[0])
	\right].
	\label{eq:reference-population-measure}
\end{align}

For each \(m\in\{0,\ldots,M\}\), define
\begin{align}
	\Delta_{n,m}
	&:=
	\widehat m_{n,K_m(n)}
	\left(
	x^t,d^t[m],\widehat\mu_{n,K_0(n)}
	\right)
	-
	\widehat m_{n,K_m(n)}
	\left(
	x^t,d^t[m],\mu_0
	\right).
	\label{eq:oracle-moment-difference}
\end{align}
Recall that the aggregation rule \(H\) satisfies, for some finite constant
\(C\),
\begin{align}
	\left|
	H(a_0,\ldots,a_M)-H(b_0,\ldots,b_M)
	\right|
	\leq
	C\max_{0\leq m\leq M}
	\left|
	\|a_m\|_2^2-\|b_m\|_2^2
	\right|.
	\label{eq:H-lipschitz-recall}
\end{align}
Here and throughout, $||.||_{2}$ denotes the Euclidean norm. 

Since $\left|
\|a+b\|_2^2-\|a\|_2^2
\right|
\leq
2\|a\|_2\|b\|_2+\|b\|_2^2$, it follows that
\begin{align}
	\left|
	\widehat T_{n,\mathbf K(n)}
	-
	T^0_{n,\mathbf K(n)}
	\right|
	\leq
	C\max_{0\leq m\leq M}
	\Bigg\{
	&
	2
	\left\|
	\widehat m_{n,K_m(n)}
	\left(x^t,d^t[m],\mu_0\right)
	\right\|_2
	\|\Delta_{n,m}\|_2 +
	\|\Delta_{n,m}\|_2^2
	\Bigg\}.
	\label{eq:oracle-statistic-decomposition}
\end{align}
Thus, it suffices to control the magnitude of the oracle moments and the
perturbations \(\Delta_{n,m}\).

For each \(m\in\{0,\ldots,M\}\), recall $Q(\cdot \mid x^{t},d^{t}[m])  : = P\left(
Y^{t-1}\in\cdot
\,\middle|\,
(X^t,D^t)=(x^t,d^t[m])
\right)$ and \begin{align}
	F_m(a\mid y^{t-1})
	&:=
	P\left(
	Y_t\leq a
	\,\middle|\,
	Y^{t-1}=y^{t-1},
	(X^t,D^t)=(x^t,d^t[m])
	\right).
	\label{eq:profile-cdf}
\end{align}
Suppose that there exist \(\delta>0\) and a $L^{2}$ function
\(A\) such that
\begin{align}
	\left|
	F_m(a\mid y^{t-1})-F_m(b\mid y^{t-1})
	\right|
	\leq
	A(y^{t-1})|a-b|^\delta
	\label{eq:profile-cdf-holder}
\end{align}
for every \(m\in\{0,\ldots,M\}\), every \(a,b\in\mathbb R\), and almost every
\(y^{t-1}\).

Let \(q\in[1,\infty]\), and let \(q'\) denote its conjugate exponent, with
\(1/q+1/q'=1\). Suppose that, for some sequence \(r_n\downarrow0\),
\begin{align}
	\max_{0\leq m\leq M}
	\left\|
	\widehat\mu_{n,K_0(n)}
	\left(\cdot,x^t,0^t,\tau\right)
	-
	\mu_0
	\right\|_{L^{\delta q}(Q(\cdot \mid x^{t},d^{t}[m]) )}
	=
	O_P(r_n).
	\label{eq:reference-estimator-rate}
\end{align}
Condition~\eqref{eq:reference-estimator-rate} requires the reference-profile
estimator to converge to \(\mu_0\) under the distributions of \(Y^{t-1}\)
prevailing in all the profile groups in which it will be evaluated.

Define
\begin{align}
	a_n
	&:=
	r_n^\delta
	\max_{0\leq m\leq M}
	\left\|
	A\|\Phi_{K_m(n)}\|_2
	\right\|_{L^{q'}( Q(\cdot \mid x^{t},d^{t}[m])  )},
	\label{eq:oracle-an}\\
	b_n
	&:=
	\max_{0\leq m\leq M}
	\left\|
	\|\Phi_{K_m(n)}\|_2
	\right\|_{L^2(Q(\cdot \mid x^{t},d^{t}[m])  )}.
	\label{eq:oracle-bn}
\end{align}
The sequence \(a_n\) controls the population effect of replacing \(\mu_0\)
with its estimator in the profile-specific moments. Indeed, the H\"older
condition and H\"older's inequality imply that this effect is bounded by a
constant multiple of $r_n^\delta
\left\|
A\|\Phi_{K_m(n)}\|_2
\right\|_{L^{q'}(Q(\cdot \mid x^{t},d^{t}[m]) )}$. 
The sequence \(b_n\) controls the magnitude of the oracle moments.

An additional condition is needed if the reference-profile moment is included. Consider the vector-valued class
\begin{align}
	\mathcal M_{n,0}
	:=
	\Big\{
	&y^{t}
	\mapsto
	\left[
	\rho\bigl(y_t,h(y^{t-1})\bigr)
	-
	\rho\bigl(y_t,\mu_0(y^{t-1})\bigr)
	\right]
	\Phi_{K_0(n)}(y^{t-1}) \colon	h\in\mathcal H_{K_0(n)}(0^t)
	\Big\}.
	\label{eq:reference-moment-class}
\end{align}
Suppose that, for some sequence \(s_n\downarrow0\),
\begin{align}
	\sup_{f\in\mathcal M_{n,0}}
	\left\|
	(P_{n,0}-P_0)f
	\right\|_2
	=
	O_P(s_n).
	\label{eq:reference-ulln}
\end{align}
Finally, define
\begin{align}
	\widetilde a_n:=a_n+s_n.
	\label{eq:oracle-atilde}
\end{align}

The preceding conditions imply
\begin{align}
	\max_{0\leq m\leq M}
	\|\Delta_{n,m}\|_2
	&=
	O_P(\widetilde a_n),
	\label{eq:moment-perturbation-rate}\\
	\max_{0\leq m\leq M}
	\left\|
	\widehat m_{n,K_m(n)}
	\left(x^t,d^t[m],\mu_0\right)
	\right\|_2
	&=
	O_P(b_n).
	\label{eq:oracle-moment-rate}
\end{align}
Combining these bounds with
\eqref{eq:oracle-statistic-decomposition} yields the following result.

\begin{proposition}[Sufficient conditions for oracle approximation]
	\label{prop:primitive-oracle-approximation}
	Suppose Assumption~1 and conditions
	\eqref{eq:profile-cdf-holder},
	\eqref{eq:reference-estimator-rate}, and
	\eqref{eq:reference-ulln} hold. If
	\begin{align}
		b_n\widetilde a_n+\widetilde a_n^2
		=
		o_{P}(\epsilon_n),
		\label{eq:primitive-oracle-rate-condition}
	\end{align}
	then
	\begin{align}
		P\left(
		\left|
		\widehat T_{n,\mathbf K(n)}
		-
		T^0_{n,\mathbf K(n)}
		\right|
		>
		\epsilon_n
		\right)
		\longrightarrow0.
		\label{eq:primitive-oracle-conclusion}
	\end{align}
	Consequently, Assumption \ref{ass:oracle-approximation} holds.
\end{proposition}

\begin{proof}
	See Appendix \ref{app:primitive-oracle-approximation}.
\end{proof}

The distinction between \(a_n\) and \(s_n\) reflects the different roles played
by the reference and nonreference profile groups. For every \(m\geq1\), the
estimator \(\widehat\mu_{n,K_0(n)}\) is constructed from the reference group and
then evaluated using observations from a different profile group. In this case,
the H\"older condition and the convergence rate in
\eqref{eq:reference-estimator-rate} are sufficient to control the resulting
moment perturbation.

For \(m=0\), by contrast, the same reference-group observations are used both to
construct \(\widehat\mu_{n,K_0(n)}\) and to evaluate the moment. The H\"older and
convergence-rate conditions continue to control the corresponding population
effect, while condition~\eqref{eq:reference-ulln} controls the difference
between that population effect and its reference-group sample counterpart.
Accordingly, condition~\eqref{eq:reference-ulln} is needed only because the
reference-profile moment is included in the test statistic. If the statistic is
constructed exclusively from the moments corresponding to
\(d^t[1],\ldots,d^t[M]\), this condition can be omitted and one may take
\(s_n=0\).

\subsection{Discussion of the Assumptions in Proposition \ref{prop:primitive-oracle-approximation}}
\label{sec:discussion-oracle-assumptions}

The conditions in Proposition~\ref{prop:primitive-oracle-approximation} have three
distinct roles. The H\"older condition translates estimation error in the
reference-profile quantile function into perturbations of the profile-specific
moments. Condition~\eqref{eq:reference-estimator-rate} controls the magnitude of
that estimation error in every profile group in which the estimator is evaluated.
Finally, condition~\eqref{eq:reference-ulln} controls the additional
empirical-process error that arises because the reference-profile estimator and
the reference-profile moment are constructed from the same observations. We
discuss each condition in turn.

\paragraph{Smoothness of the conditional distribution.}

Condition~\eqref{eq:profile-cdf-holder} controls how much the conditional
probability
\[
P\left(
Y_t\leq h(Y^{t-1})
\,\middle|\,
Y^{t-1},(X^t,D^t)=(x^t,d^t[m])
\right)
\]
changes when \(h\) is perturbed. This is the link between convergence of
\(\widehat\mu_{n,K_0(n)}\) as a function and convergence of the corresponding
quantile moments. In particular, if
\(\widehat\mu_{n,K_0(n)}-\mu_0\) has magnitude \(r_n\), the H\"older condition
implies a moment perturbation of magnitude \(r_n^\delta\).

The leading case is \(\delta=1\). For example, condition
\eqref{eq:profile-cdf-holder} holds with \(\delta=1\) if the conditional
distribution of \(Y_t\) admits a density that is locally bounded over the range
of values taken by \(\mu_0\) and its sieve approximations. Allowing
\(\delta<1\) accommodates conditional distributions that are less regular near
the quantile of interest. The function \(A\) allows the degree of smoothness to
depend on \(Y^{t-1}\). It need not be uniformly bounded, provided its interaction
with the moment vector satisfies
\[
\left\|
A\|\Phi_{K_m(n)}\|_2
\right\|_{L^{q'}(Q(\cdot \mid x^{t},d^{t}[m]) )}
<\infty.
\]
Thus, the condition allows the conditional density, or the corresponding
H\"older constant, to vary with the conditioning variables.

\paragraph{Convergence of the reference-profile estimator.}

Condition~\eqref{eq:reference-estimator-rate} requires a convergence
rate for the estimator constructed from the reference group. Because
the reference-group size grows proportionally with \(n\), this rate can
be obtained from standard results, both when the reference-profile
quantile function is parametrically specified and when it is estimated
semiparametrically or nonparametrically.

The relevant distance is an \(L^{\delta q}\) distance because the
H\"older condition raises the estimation error to the power \(\delta\).
Indeed, $\left\|
|\widehat\mu_{n,K_0(n)}-\mu_0|^\delta
\right\|_{L^q(Q(\cdot \mid x^{t},d^{t}[m])  )}
=
\left\|
\widehat\mu_{n,K_0(n)}-\mu_0
\right\|_{L^{\delta q}(Q(\cdot \mid x^{t},d^{t}[m]) )}^\delta$. 
Thus, a functional estimation rate \(r_n\) produces a population
moment error of order \(r_n^\delta\).

Parametric and semi/nonparametric cases are discussed below, but now we point out that in both cases, convergence
is required under the distributions \(P_m\) prevailing in all the
profile groups, rather than only under the reference-group distribution.
This is necessary because the estimator constructed from the
reference group is subsequently evaluated using observations from
every profile group included in the statistic.

Rates established in more conventional norms can nevertheless be
used to verify this requirement. Most directly, $\max_{0\leq m\leq M}
\|h\|_{L^{\delta q}(Q(\cdot \mid x^{t},d^{t}[m]) )}
\leq
\|h\|_{L^\infty}$. 
Alternatively, if, for some finite \(C\), $P\left(
(X^t,D^t)=(x^t,d^t[m])
\right)
\geq C^{-1}$ 
for every \(m\in\{0,\ldots,M\}\), then $\max_{0\leq m\leq M}
\|h\|_{L^{\delta q}(Q(\cdot \mid x^{t},d^{t}[m])  )}
\leq
C^{1/(\delta q)}
\|h\|_{L^{\delta q}(P)}$. 
Thus, a rate obtained under the unconditional distribution also
verifies condition~\eqref{eq:reference-estimator-rate}. More
generally, the same conclusion follows whenever the profile-specific
distributions are uniformly dominated by a distribution under which
the relevant parametric or sieve-estimation rate is available.

\paragraph{Growth of the moment dimensions.}

The sequences \(a_n\) and \(b_n\) summarize the restrictions on the dimensions
\(K_0(n),\ldots,K_M(n)\). The quantity
\[
a_n
=
r_n^\delta
\max_{0\leq m\leq M}
\left\|
A\|\Phi_{K_m(n)}\|_2
\right\|_{L^{q'}(Q(\cdot \mid x^{t},d^{t}[m]) )}
\]
combines the convergence rate of the reference-profile estimator with the
magnitude of the profile-specific moment vectors. The quantity
\[
b_n
=
\max_{0\leq m\leq M}
\left\|
\|\Phi_{K_m(n)}\|_2
\right\|_{L^2(Q(\cdot \mid x^{t},d^{t}[m]) )}
\]
controls the magnitude of the oracle moments themselves. Consequently, faster
growth of the moment dimensions increases both the direct effect of estimating
\(\mu_0\) and the interaction between that estimation error and the oracle
moments.

In the setting emphasized in this paper, some treatment-profile group sizes may
remain bounded. There is then generally no reason to let the corresponding
moment dimensions \(K_m(n)\), \(m\geq1\), grow with \(n\). These dimensions can
be held fixed without restricting the flexibility of an estimator that could
not, in any event, consistently estimate an increasing number of moments from a
bounded number of observations.

If all the moment dimensions are fixed and the required envelope moments are
bounded, then $b_n=O(1)$ and $a_n=O(r_n^\delta)$. In this case, the rate condition in
\eqref{eq:primitive-oracle-rate-condition} reduces to $r_n^\delta+s_n=o(\epsilon_n)$. 
Thus, \(\epsilon_n\) can be chosen to converge to zero sufficiently slowly
relative to the estimation and empirical-process errors. When some of the
moment dimensions grow, condition
\eqref{eq:primitive-oracle-rate-condition} instead determines how quickly they
may grow relative to \(r_n\), \(s_n\), and \(\epsilon_n\).

The two terms in $b_n\widetilde a_n+\widetilde a_n^2$ have a direct interpretation. The first is the interaction between the oracle
moment and the perturbation generated by estimating \(\mu_0\). The second is the
quadratic contribution of that perturbation. Both terms arise because the test
statistic aggregates squared norms of the profile-specific moments.

\paragraph{The reference-group empirical-process condition.}

Condition~\eqref{eq:reference-ulln} is a uniform law of large numbers over the
sieve class used to estimate the reference-profile conditional quantile
function. It controls the complexity of
\(\mathcal H_{K_0(n)}(0^t)\), together with the dimension and magnitude of
\(\Phi_{K_0(n)}\). It can be established using standard entropy conditions for
classes indexed by conditional-quantile functions.

This condition is needed because \(d^t[0]\) is included among the profiles
entering the test statistic. In the reference group, the same observations are
used both to construct \(\widehat\mu_{n,K_0(n)}\) and to evaluate the sample
moment. Therefore, the H\"older and convergence-rate conditions alone control
only the population effect of replacing \(\mu_0\) by its estimator. Condition
\eqref{eq:reference-ulln} controls the difference between this population effect
and its reference-group sample counterpart.

By contrast, for \(m\geq1\), the reference-profile estimator is evaluated using
observations outside the group from which it was constructed. Assumption~1 then
allows the resulting perturbation to be controlled directly through its
conditional first moment, even if the number of observations in that profile
group does not diverge. No profile-specific uniform law of large numbers is
required for these groups.

Condition~\eqref{eq:reference-ulln} can be omitted if the reference-profile
moment is excluded from the test statistic. In that case, one may set \(s_n=0\),
and the oracle-approximation condition depends only on the population effect
\(a_n\). Primitive entropy conditions implying
\eqref{eq:reference-ulln} when the reference-profile moment is included are
provided below.

We conclude by providing primitive conditions for the reference-group
uniform law in \eqref{eq:reference-ulln}. The result specializes the standard
bracketing argument for indicator-indexed classes (e.g. \cite{CvKL2003}) to the
instrument-weighted, vector-valued class \(\mathcal M_{n,0}\).

Maintain the conditions
imposed above, and suppose that
\begin{align}
	\mathcal H_{K_0(n)}(0^t)
	\subseteq
	\mathcal H
	\qquad
	\text{for every }n,
	\label{eq:reference-sieve-containment}
\end{align}
where \(\mathcal H\) is a uniformly bounded H\"older ball of functions
on a compact set \(\mathcal Y\subseteq\mathbb R^{d_y}\), with
smoothness exponent \(\upsilon\). Define
\begin{align}
	c_{\Phi,n}
	&:=
	\left\|
	\|\Phi_{K_0(n)}\|_2
	\right\|_{L^\infty}.
	\label{eq:reference-instrument-envelope}
\end{align}

The following result gives sufficient conditions for
\eqref{eq:reference-ulln}.

\begin{proposition}
	\label{prop:primitive-reference-ulln}
	Suppose Assumption~1 and the conditions imposed above hold. If $\upsilon>\frac{d_y}{\delta}$, 
	then
	\begin{align}
		\sup_{f\in\mathcal M_{n,0}}
		\left\|
		(P_{n,0}-P_0)f
		\right\|_2
		=
		O_P\left(
		\frac{c_{\Phi,n}}{\sqrt n}
		\right).
		\label{eq:primitive-reference-ulln-rate}
	\end{align}
	Consequently, if $\frac{c_{\Phi,n}}{\sqrt n}\rightarrow0$, condition~\eqref{eq:reference-ulln} holds with $s_n
	=
	\frac{c_{\Phi,n}}{\sqrt n}$. 
\end{proposition}

\begin{proof}
	See Appendix \ref{app:primitive-oracle-approximation}. 
\end{proof}

Thus, the primitive condition separates the complexity of the
reference-profile function class from the magnitude of the instrument
vector. The H\"older restriction $\upsilon>\frac{d_y}{\delta}$ ensures that the indicator class
has a finite entropy integral, while \(c_{\Phi,n}\) determines how the
dimension and magnitude of the reference-group moments affect the
uniform-law rate.

\subsubsection{Comments for Parametric and Semi/Nonparametric Specifications}

\paragraph{Parametric specification.} Suppose that the reference-profile quantile function is indexed by a
finite-dimensional parameter: $\mu_0(y^{t-1})
=
\mu(y^{t-1};\beta_0)$, for $\beta_0\in\mathbb R^{d_\beta}$. 
Under the standard regularity conditions for parametric quantile
regression, the reference-group estimator satisfies $\|\widehat\beta_n-\beta_0\|_2
= O_P(n^{-1/2})$;  see, for example, \cite{koenker1978regression}. If the parametrization is
locally Lipschitz, in the sense that
\[
\left|
\mu(y^{t-1};\beta)-\mu(y^{t-1};\beta_0)
\right|
\leq
L(y^{t-1})\|\beta-\beta_0\|_2
\]
for \(\beta\) in a neighborhood of \(\beta_0\), and $\max_{0\leq m\leq M}
\|L\|_{L^{\delta q}(Q(\cdot \mid x^{t},d^{t}[m]) )}
<\infty$, 
then $\max_{0\leq m\leq M}
\left\|
\mu(\cdot;\widehat\beta_n)-\mu(\cdot;\beta_0)
\right\|_{L^{\delta q}(Q(\cdot \mid x^{t},d^{t}[m]) )}
=
O_{P}(n^{-1/2})$ and condition~\eqref{eq:reference-estimator-rate} holds with $r_n=n^{-1/2}$. 
The corresponding population moment error is therefore of order
\(n^{-\delta/2}\), and of the usual root-\(n\) order when
\(\delta=1\).

In parametric cases, $K(n)$ is not required to grow, so $b_{n} = O(1)$ and $a_{n} = O(n^{-\delta/2})$ and $c_{\Phi,n} = O(1)$ so that $s_{n} = n^{-1/2}$. Hence, $\epsilon_{n}$ can be chosen to be 
\begin{align}
	\epsilon_{n} =  \ell_{n}  n^{-\delta/2}  
\end{align}
where $(\ell_{n})_{n}$ is an arbitrarily  slowly diverging sequence; e.g. $l_{n} = \log \log n$.

\paragraph{Semi-/Non-parametric specifications.} When \(\mu_0\) is infinite-dimensional, the estimator
\(\widehat\mu_{n,K_0(n)}\) is a sieve minimum-distance estimator based
on the reference group. Since the reference-group size is proportional
to \(n\), available results for semiparametric and nonparametric
conditional-moment models apply; e.g., see \cite{chen2007large,chen2009efficient,horowitz2007nonparametric,gagliardini2012nonparametric}.

For illustration, suppose that \(\mu_0\) is \(s\)-smooth, its arguments
have dimension \(d\), and \(J_n\) denotes the dimension of the sieve
used to estimate it. In a well-posed model, a standard rate, up to
logarithmic or arbitrarily slowly diverging factors, is $r_n
\asymp
J_n^{-s/d}
+
\sqrt{\frac{J_n}{n}}$. 
The first term is the sieve approximation error and the second is the
sampling error. Choosing $J_n\asymp n^{d/(2s+d)}$  gives $r_n
\asymp
n^{-s/(2s+d)}$. 
In an ill-posed conditional-moment model, the sampling component is
multiplied by the relevant sieve measure of ill-posedness,
\(\tau_{J_n}\), so that $r_n
\asymp
J_n^{-s/d}
+
\tau_{J_n}\sqrt{\frac{J_n}{n}}$. 
Accordingly, the rate depends on both the approximation properties of
the sieve and the degree of ill-posedness. The results in the cited
literature commonly yield rates of the form
\[
r_n=n^{- \delta \iota}\ell_n,
\qquad
\iota\in(0,1/2),
\]
where \(\ell_n\) is logarithmic or arbitrarily slowly diverging and
\(\iota\) depends on smoothness, sieve approximation, and
ill-posedness.

The dimensions
\(K_m(n)\) determine the numbers of moments entering
the test statistic. In the case where the treatment groups \emph{do not grow} with the sample size, $K_{m}(n)$ can remain bounded for all $m \geq 1$. However, $K_{0}(n)$ will have to grow. 

Thus, for the well-posed case with the rate-optimal choice for instance, $\epsilon_{n}$ will satisfy
\begin{align*}
	b_n
	\left(
	n^{-s\delta/(2s+d)}c_{1n}
	+
	c_{\Phi,n} n^{-1/2}c_{0n}
	\right)
	+
	\left(
	n^{-s\delta/(2s+d)}c_{1n}
	+
n^{-1/2}c_{0n}
	\right)^2
	=
	o(\epsilon_n).
\end{align*}
where $c_{1n}
	:=
	\max_{0\leq m\leq M}
	\left\|
	A\|\Phi_{K_m(n)}\|_2
	\right\|_{L^{q'}(Q(\cdot \mid x^{t},d^{t}[m]) )}$, and $c_{0n}
	:=
	\left\|
	\|\Phi_{K_0(n)}\|_2
	\right\|_{L^\infty}$.

\subsection{Proofs of Propositions \ref{prop:primitive-oracle-approximation} and \ref{prop:primitive-reference-ulln}}
\label{app:primitive-oracle-approximation}

\begin{proof}[Proof of Proposition \ref{prop:primitive-oracle-approximation}]
	For notational simplicity, write $\widehat\mu_n
	:=
	\widehat\mu_{n,K_0(n)}
	\left(\cdot,x^t,0^t,\tau\right)$, $\Phi_{i,m}
	:=
	\Phi_{K_m(n)}(Y_i^{t-1})$, and let $R_{n,m}
	:=
	\|\widehat\mu_n-\mu_0\|_{L^{\delta q}(Q(\cdot \mid x^{t},d^{t}[m]) )}$. 
	By condition~\eqref{eq:reference-estimator-rate},
	\begin{align}
		\max_{0\leq m\leq M}R_{n,m}
		=
		O_P(r_n).
		\label{eq:proof-reference-rate}
	\end{align}
	
	We first control the moment perturbations \(\Delta_{n,m}\). The argument differs
	slightly between the nonreference and reference profile groups.
	
	\paragraph{Step 1: Nonreference profile groups.}
	
	Fix \(m\in\{1,\ldots,M\}\). By the definition of the sample moments,
	\begin{align}
		\Delta_{n,m}
		=
		\frac{1}{N_n((x^t,d^t[m]))}
		\sum_{i\in\mathcal G_n((x^t,d^t[m]))}
		\Big[
		\rho\bigl(Y_{i,t},\widehat\mu_n(Y_i^{t-1})\bigr)
		-
		\rho\bigl(Y_{i,t},\mu_0(Y_i^{t-1})\bigr)
		\Big]
		\Phi_{i,m}.
		\label{eq:proof-delta-treated}
	\end{align}
	
	For any measurable function \(h\), the definition of the quantile score gives
	\begin{align}
		&
		E\left[
		\left|
		\rho\bigl(Y_t,h(Y^{t-1})\bigr)
		-
		\rho\bigl(Y_t,\mu_0(Y^{t-1})\bigr)
		\right|
		\,\middle|\,
		Y^{t-1}=y^{t-1},
		(X^t,D^t)=(x^t,d^t[m])
		\right]
		\nonumber\\
		&\hspace{2cm}
		=
		\left|
		F_m\bigl(h(y^{t-1})\mid y^{t-1}\bigr)
		-
		F_m\bigl(\mu_0(y^{t-1})\mid y^{t-1}\bigr)
		\right|.
		\label{eq:proof-indicator-difference}
	\end{align}
	
	Because \(\widehat\mu_n\) is constructed using observations from the reference
	group, Assumption~1 implies that, conditionally on the reference-group observations
	and the profile-group memberships, the observations in group \(m\) are independent
	of \(\widehat\mu_n\) and have the profile-\(m\) conditional distribution. Thus, conditioning on $	\widehat\mu_n$ does not affect the distribution of $\Delta_{n,m}$. This observation, the previous one, and the triangle inequality imply
	\begin{align}
		E\left[
		\|\Delta_{n,m}\|_2
		\,\middle|\,
		\widehat\mu_n
		\right]
		&\leq
		E_{P_m}\left[
		\left|
		F_m\bigl(\widehat\mu_n(Y^{t-1})\mid Y^{t-1}\bigr)
		-
		F_m\bigl(\mu_0(Y^{t-1})\mid Y^{t-1}\bigr)
		\right|
		\|\Phi_{K_m(n)}(Y^{t-1})\|_2
		\right]
		\nonumber\\
		&\leq
		E_{P_m}\left[
		A(Y^{t-1})
		|\widehat\mu_n(Y^{t-1})-\mu_0(Y^{t-1})|^\delta
		\|\Phi_{K_m(n)}(Y^{t-1})\|_2
		\right],
		\label{eq:proof-holder-first}
	\end{align}
	where the second inequality follows from
	\eqref{eq:profile-cdf-holder}. H\"older's inequality now yields
	\begin{align}
		E\left[
		\|\Delta_{n,m}\|_2
		\,\middle|\,
		\widehat\mu_n
		\right]
		&\leq
		\left\|
		|\widehat\mu_n-\mu_0|^\delta
		\right\|_{L^q(Q(\cdot \mid x^{t},d^{t}[m]) )}
		\left\|
		A\|\Phi_{K_m(n)}\|_2
		\right\|_{L^{q'}(Q(\cdot \mid x^{t},d^{t}[m]) )}
		\nonumber\\
		&=
		R_{n,m}^\delta
		\left\|
		A\|\Phi_{K_m(n)}\|_2
		\right\|_{L^{q'}(Q(\cdot \mid x^{t},d^{t}[m]) )}.
		\label{eq:proof-holder-bound}
	\end{align}
	Together with \eqref{eq:proof-reference-rate} and the definition of \(a_n\),
	this implies
	\begin{align}
		\|\Delta_{n,m}\|_2
		=
		O_P(a_n),
		\qquad
		m\in\{1,\ldots,M\}.
		\label{eq:proof-treated-delta-rate}
	\end{align}
	Notice that this argument uses only a first-moment bound and therefore does not
	require \(N_n(x^t,d^t[m])\) to diverge.
	
	\paragraph{Step 2: The reference-profile group.}
	
	For the reference group, define
	\[
	f_{\widehat\mu_n}(y_t,y^{t-1})
	:=
	\Big[
	\rho\bigl(y_t,\widehat\mu_n(y^{t-1})\bigr)
	-
	\rho\bigl(y_t,\mu_0(y^{t-1})\bigr)
	\Big]
	\Phi_{K_0(n)}(y^{t-1}).
	\]
	Since \(\widehat\mu_n\in\mathcal H_{K_0(n)}(0^t)\) by construction,
	\(f_{\widehat\mu_n}\in\mathcal M_{n,0}\). Moreover, $\Delta_{n,0}=P_{n,0}f_{\widehat\mu_n}$. 
	Adding and subtracting its population counterpart gives
	\begin{align}
		\|\Delta_{n,0}\|_2
		\leq
		\left\|
		(P_{n,0}-P_0)f_{\widehat\mu_n}
		\right\|_2
		+
		\left\|
		P_0f_{\widehat\mu_n}
		\right\|_2.
		\label{eq:proof-reference-decomposition}
	\end{align}
	By condition~\eqref{eq:reference-ulln},
	\begin{align}
		\left\|
		(P_{n,0}-P_0)f_{\widehat\mu_n}
		\right\|_2
		&\leq
		\sup_{f\in\mathcal M_{n,0}}
		\|(P_{n,0}-P_0)f\|_2
		=
		O_P(s_n).
		\label{eq:proof-reference-empirical}
	\end{align}
	
	For the population term, the same argument used in
	\eqref{eq:proof-indicator-difference}--\eqref{eq:proof-holder-bound} gives
	\begin{align}
		\left\|
		P_0f_{\widehat\mu_n}
		\right\|_2
		&\leq
		E_{P_0}\left[
		A(Y^{t-1})
		|\widehat\mu_n(Y^{t-1})-\mu_0(Y^{t-1})|^\delta
		\|\Phi_{K_0(n)}(Y^{t-1})\|_2
		\right]
		\nonumber\\
		&\leq
		R_{n,0}^\delta
		\left\|
		A\|\Phi_{K_0(n)}\|_2
		\right\|_{L^{q'}(Q(\cdot \mid x^{t},d^{t}[0]) )}
		=
		O_P(a_n).
		\label{eq:proof-reference-population}
	\end{align}
	Consequently,
	\begin{align}
		\|\Delta_{n,0}\|_2
		=
		O_P(a_n+s_n)
		=
		O_P(\widetilde a_n).
		\label{eq:proof-reference-delta-rate}
	\end{align}
	Since the number of profile groups is finite, combining
	\eqref{eq:proof-treated-delta-rate} and
	\eqref{eq:proof-reference-delta-rate} yields
	\begin{align}
		\max_{0\leq m\leq M}\|\Delta_{n,m}\|_2
		=
		O_P(\widetilde a_n).
		\label{eq:proof-all-delta-rate}
	\end{align}
	
	\paragraph{Step 3: Magnitude of the oracle moments.}
	
	For every \(m\in\{0,\ldots,M\}\),
	\begin{align*}
		\left\|
		\widehat m_{n,K_m(n)}
		\left(x^t,d^t[m],\mu_0\right)
		\right\|_2
		&\leq
		\frac{1}{N_n(x^t,d^t[m])}
		\sum_{i\in\mathcal G_n(x^t,d^t[m])}
		\left|
		\rho\bigl(Y_{i,t},\mu_0(Y_i^{t-1})\bigr) - \tau
		\right|
		\|\Phi_{i,m}\|_2
		\nonumber\\
		&\leq
		\frac{1}{N_n(x^t,d^t[m])}
		\sum_{i\in\mathcal G_n(x^t,d^t[m])}
		\|\Phi_{i,m}\|_2,
	\end{align*}
	because \(|\rho(y,a)|\leq1\). Conditional on the profile-group memberships, the
	expectation of the right-hand side is bounded by $E_{Q(\cdot \mid x^{t},d^{t}[m]) }\left[
		\|\Phi_{K_m(n)}(Y^{t-1})\|_2
		\right]
		\leq
		\left\|
		\|\Phi_{K_m(n)}\|_2
		\right\|_{L^2(Q(\cdot \mid x^{t},d^{t}[m]) )}
		\leq b_n$. 
	Markov's inequality and finiteness of \(M\) therefore imply
	\begin{align}
		\max_{0\leq m\leq M}
		\left\|
		\widehat m_{n,K_m(n)}
		\left(x^t,d^t[m],\mu_0\right)
		\right\|_2
		=
		O_P(b_n).
		\label{eq:proof-all-oracle-moments}
	\end{align}
	This bound also does not require the nonreference profile-group sizes to diverge.
	
	\paragraph{Step 4: Approximation of the test statistic.}
	
	Applying \eqref{eq:proof-all-delta-rate} and
	\eqref{eq:proof-all-oracle-moments} to
	\eqref{eq:oracle-statistic-decomposition} gives
	\begin{align}
		\left|
		\widehat T_{n,\mathbf K(n)}
		-
		T^0_{n,\mathbf K(n)}
		\right|
		=
		O_P\left(
		b_n\widetilde a_n+\widetilde a_n^2
		\right).
		\label{eq:proof-final-oracle-rate}
	\end{align}
	By condition~\eqref{eq:primitive-oracle-rate-condition}, $b_n\widetilde a_n+\widetilde a_n^2=o(\epsilon_n)$. 
	It follows that $\frac{
		\left|
		\widehat T_{n,\mathbf K(n)}
		-
		T^0_{n,\mathbf K(n)}
		\right|
	}{
		\epsilon_n
	}
	\overset{P}{\longrightarrow}0$, 
	and hence
	\[
	P\left(
	\left|
	\widehat T_{n,\mathbf K(n)}
	-
	T^0_{n,\mathbf K(n)}
	\right|
	>
	\epsilon_n
	\right)
	\longrightarrow0.
	\]
	This is precisely the oracle-approximation requirement in Assumption~4.
\end{proof}

\begin{proof}[Proof of Proposition \ref{prop:primitive-reference-ulln}]
	For \(h\in\mathcal H\), define
	\begin{align}
		a_h(y_t,y^{t-1})
		&:=
		\rho\bigl(y_t,h(y^{t-1})\bigr)
		-
		\rho\bigl(y_t,\mu_0(y^{t-1})\bigr),
		\label{eq:proof-reference-ah}\\
		f_{n,h}(y_t,y^{t-1})
		&:=
		a_h(y_t,y^{t-1})
		\Phi_{K_0(n)}(y^{t-1}).
		\label{eq:proof-reference-fnh}
	\end{align}
	Because of \eqref{eq:reference-sieve-containment},  $\mathcal M_{n,0}
	\subseteq
	\mathcal F_n
	:=
	\{f_{n,h}:h\in\mathcal H\}$. 
	It therefore suffices to establish the desired bound uniformly over
	\(\mathcal F_n\).
	
	Consider first the scalar indicator class $\mathcal I
	:=
	\left\{
	(y_t,y^{t-1})
	\mapsto
	\rho\bigl(y_t,h(y^{t-1})\bigr)
	:
	h\in\mathcal H
	\right\}$. 
	The standard entropy bound for a H\"older ball gives
	\begin{align}
		\log N\left(
		\eta,
		\mathcal H,
		\|\cdot\|_\infty
		\right)
		\lesssim
		\eta^{-d_y/\upsilon}, \quad 	\eta\downarrow0.
		\label{eq:proof-holder-covering-number}
	\end{align}

	Let \(\{h_l:1\leq l\leq L(\eta)\}\) be an
	\(\eta\)-cover of \(\mathcal H\) under the uniform norm. For any
	\(h\in\mathcal H\), there exists \(h_l\) such that $h_l-\eta\leq h\leq h_l+\eta$. 
	Consequently,
	\begin{align}
		\rho\bigl(y_t,h_l(y^{t-1})-\eta\bigr)
		\leq
		\rho\bigl(y_t,h(y^{t-1})\bigr)
		\leq
		\rho\bigl(y_t,h_l(y^{t-1})+\eta\bigr).
		\label{eq:proof-reference-indicator-bracket}
	\end{align}
	By the H\"older-continuity condition imposed above, the squared
	\(L^2( P_0 )\) width of this bracket is bounded by
	\begin{align}
		&E_{P_0 }\left[
		\left|
		\rho\bigl(Y_t,h_l(Y^{t-1})+\eta\bigr)
		-
		\rho\bigl(Y_t,h_l(Y^{t-1})-\eta\bigr)
		\right|^2
		\right]
		\nonumber\\
		&\quad=
		E_{Q_0(\cdot \mid X^{t},D^{t} = (x^{t},d^{t}[0]))}\left[
		F_0\bigl(h_l(Y^{t-1})+\eta\mid Y^{t-1}\bigr)
		-
		F_0\bigl(h_l(Y^{t-1})-\eta\mid Y^{t-1}\bigr)
		\right]
		\lesssim
		\eta^\delta.
		\label{eq:proof-reference-bracket-width}
	\end{align}
	An \(L^2( P_0  )\) bracket of width \(\varepsilon\) can therefore be
	obtained by taking
	\(\eta\asymp\varepsilon^{2/\delta}\). It follows from
	\eqref{eq:proof-holder-covering-number} that $\log N_{[]}\left(
	\varepsilon,
	\mathcal I,
	L^2( P_0 )
	\right)
	\lesssim
	\varepsilon^{-2d_y/(\delta\upsilon)}$. 
	Hence,
	\begin{align}
		\int_0^1
		\sqrt{
			\log N_{[]}\left(
			\varepsilon,
			\mathcal I,
			L^2( P_0 )
			\right)
		}
		\,d\varepsilon
		<\infty
		\label{eq:proof-reference-entropy-integral}
	\end{align}
	whenever \(\upsilon>d_y/\delta\).
	
	Subtracting the fixed function
	\(\rho(y_t,\mu_0(y^{t-1}))\) does not alter the entropy bound. Thus, the
	class $\mathcal A:=\{a_h:h\in\mathcal H\} $ with $a_h(y_t,y^{t-1})
	=
1\{y_t\le h(y^{t-1})\}
	-
1\{y_t\le \mu_0(y^{t-1})\}$ also has a finite \(L^2(P_0 )\) bracketing integral.
	
	We next incorporate the vector of instruments. If \(c_{\Phi,n}>0\),
	write $\Psi_n
	:=
	\frac{\Phi_{K_0(n)}}{c_{\Phi,n}}$. 
	By construction, $\left\|
	\|\Psi_n\|_2
	\right\|_{L^\infty}
	\leq1$. 
	Consequently, for any \(h,\widetilde h\in\mathcal H\), $\left\|
	(a_h-a_{\widetilde h})\Psi_n
	\right\|_{L^2(P_0 ;\ell_2)}
	\leq
	\left\|
	a_h-a_{\widetilde h}
	\right\|_{L^{2}(P_0)}$. The normalized vector-valued class	$\widetilde{\mathcal F}_n
	:=
	\left\{
	a_h\Psi_n:h\in\mathcal H
	\right\}$ 
	therefore inherits the entropy bound in
	\eqref{eq:proof-reference-entropy-integral}. The standard
	Hilbert-valued empirical-process maximal inequality, together with
	\eqref{eq:proof-reference-entropy-integral}, gives
	\begin{align}
		\sup_{h\in\mathcal H}
		\left\|
		(P_n-P)
		\left[
		1\{G_i^t=g_0^t\}f_{n,h}
		\right]
		\right\|_2
		=
		O_P\left(
		\frac{c_{\Phi,n}}{\sqrt n}
		\right).
		\label{eq:proof-reference-unconditional-process}
	\end{align}
	
	It remains to account for the reference-group normalization. Let $\widehat p_{0,n}
	:=
	P_n1\{G_i^t=g_0^t\}
	=
	\frac{N_n(g_0^t)}{n}$.  For every \(h\in\mathcal H\),
	\begin{align}
		(P_{n,0}-P_0)f_{n,h}
		=
		\frac{1}{\widehat p_{0,n}}
		\Bigg\{
		&(P_n-P)
		\left[
		1\{G_i^t=g_0^t\}f_{n,h}
		\right]-
		P_0f_{n,h}
		(P_n-P)1\{G_i^t=g_0^t\}
		\Bigg\}.
		\label{eq:proof-reference-ratio}
	\end{align}
	The conditions imposed above imply $\widehat p_{0,n}^{-1}=O_P(1)$ and  $(P_n-P)1\{G_i^t=g_0^t\}=O_P(n^{-1/2})$.  Moreover, since \(|a_h|\leq1\),
	\begin{align}
		\sup_{h\in\mathcal H}
		\|P_0f_{n,h}\|_2
		&\leq
		E_{P_0}\left[
		\|\Phi_{K_0(n)}(Y^{t-1})\|_2
		\right]
		\leq c_{\Phi,n}.
		\label{eq:proof-reference-population-term}
	\end{align}
	Combining
	\eqref{eq:proof-reference-unconditional-process}--\eqref{eq:proof-reference-population-term} with
	\eqref{eq:proof-reference-ratio} gives $\sup_{h\in\mathcal H}
		\left\|
		(P_{n,0}-P_0)f_{n,h}
		\right\|_2
		=
		O_P\left(
		\frac{c_{\Phi,n}}{\sqrt n}
		\right)$. 
	Restricting the supremum to
	\(\mathcal H_{K_0(n)}(0^t)\) establishes
	\eqref{eq:primitive-reference-ulln-rate}. The remaining conclusions
	follow immediately.
\end{proof}

\end{document}